\documentclass[a4paper,twoside,11pt]{article}

\usepackage[USenglish]{babel} 
\usepackage[T1]{fontenc}

\usepackage{amsmath}
\usepackage{graphicx}
\usepackage{url}
\usepackage[labelsep=endash, figurename=Figure, tablename=Table, textfont=it]{caption}
\usepackage{subcaption}
\usepackage{float}  
\usepackage{hyperref, wasysym}

\usepackage{multicol}

\usepackage{multirow}

\usepackage{rotating}      
\usepackage{threeparttable}
\usepackage{makecell}      
\usepackage{booktabs}      
\usepackage{float} 

\usepackage{fancyhdr}
\usepackage[top=2.5cm, bottom=2.5cm, left=2.5cm, right=2.5cm]{geometry} 
\usepackage[dvipsnames]{pstricks}

\usepackage[round,comma,authoryear]{natbib}
\usepackage{titling}
\usepackage{booktabs}
\usepackage{cleveref}

\definecolor{darkblue}{rgb}{0,0,0.5}
\hypersetup{colorlinks=true,linkcolor=black,citecolor=darkblue,urlcolor=darkblue} 

\makeatletter
\let\ps@plain\ps@fancy
\makeatother

\begin{document}


\title{\bf Evaluating Electric Micro-Mobility Related Mode Choice Stated Preferences: A Latent Class Choice Approach}
\author{Yikang Wu\textsuperscript{1}, Mehmet Yildirimoglu\textsuperscript{1}, Zuduo Zheng\textsuperscript{1}\\[1em]
    \textsuperscript{1}The University of Queensland }
\date{\today}

\pretitle{\centering\Large}
\posttitle{\par\vspace{1ex}}

\preauthor{\centering}
\postauthor{\par\vspace{1ex}

\vspace{1ex}\it

\vspace{1ex}


}

\maketitle
\vspace{-1cm}
\noindent\rule{\textwidth}{0.5pt}\vspace{0cm}

\renewcommand{\thepage}{\arabic{page}}	

\section{ Introduction}

Electric Micro-Mobility (EMM) devices, such as e-bikes and e-scooters, have emerged as sustainable
transportation options that can help alleviate traffic congestion and support decarbonisation efforts. EMM offers a flexible, sustainable, and cost-effective alternative, reducing reliance on private vehicles or public transport, particularly for short and medium-distance travel \citep{shaheen2013public, clewlow2018micro,tiwari2019micro}. A study conducted in New Zealand \citep{fitt2019scooter} suggests that EMM could potentially replace approximately 30\% of car trips. Concurrently, shared micro-mobility is increasingly recognised as a complementary mode to public transit, providing faster and more cost-effective mobility options that contribute to enhancing urban resilience \citep{cui2024integration}. 

EMM provides mode shift opportunities by its inherent advantages compared to private cars. By serving as a versatile substitute for personal cars, EMM grants access to areas where private vehicles face challenges, including narrow roads and urban city streets\citep{milakis2020micro}. The mode shift to EMM could greatly liberate road and parking spaces owing to its small scale, thus reducing congestion. Meanwhile, EMM maintains a lower emissions footprint and offers more health benefits per passenger kilometre of travel in comparison to private vehicles, significantly reducing carbon dioxide emissions\citep{reck2022mode}. 

shared micro-mobility can complement traditional public transit services by addressing the ‘first and last mile’ challenge \citep{guo2021understanding,reck2021explaining,nikiforiadis2021analysis,merlin2021segment,wang2023travel}. With its sustainable, healthy, cost-effective, and convenient attributes, shared micro-mobility demonstrates significant potential for mode shifts from other travel modes, including exclusive reliance on public transport \citep{cui2024integration}. By overcoming the limitations of walking to reach final destinations, shared micro-mobility enhances public transport accessibility. Integrating shared micro-mobility with public transit not only encourages active travel but also increases public transport usage, ultimately attracting private car users to more sustainable modes of travel. This shift can free up road space, improving the overall efficiency of the transport system.

Many studies have explored the usage patterns of both personal micromobility and shared micromobility, particularly in relation to mode choice behaviour and factors influencing EMM adoption\citep{asgari2018stated,cao2021scooter,baek2021electric,van2021insights,van2022preferences,montes2023shared}. However, while previous research has acknowledged preference heterogeneity in micromobility adoption, it has largely focused on broad demographic or socioeconomic characteristics, often relying on basic multinomial logit models or, more recently, mixed logit models \citep{eom2023exploring,cao2021scooter,baek2021electric,curtale2021understanding,montes2023shared}. These approaches, though useful, do not fully capture the varying sensitivities of different user groups toward key attributes influencing their mode choices. Additionally, the majority of existing studies have examined either personal micromobility or shared micromobility in isolation\citep{esztergar2022assessment,cao2021scooter,baek2021electric,fishman2015factors,van2022preferences}, with limited research addressing the co-existence of these two EMM modes within the same urban mobility framework.

This study enhances the current understanding of mode choice behaviour by examining the heterogeneity among latent segments of car and public transport users when presented with EMM alternatives, including both personal and shared micromobility options. Utilising stated-preference data collected from 1,671 Brisbane residents, this study employs Latent Class Choice Models (LCCMs) to identify distinct user segments and analyse their underlying behavioural patterns in response to EMM options. Based on the findings, this study proposes targeted policy recommendations to facilitate the adoption of EMM, enhance urban mobility strategies, and ultimately contribute to a more efficient and sustainable transport network.


\section{ Literature review}
EMM adoption and usage analysis applying stated preference research is getting popular recent years in various context. Several studies have examined the factors influencing micromobility adoption, highlighting the significance of cost, travel time, safety, infrastructure availability, and integration with public transport. \citep{montes2023shared} and \citep{van2022preferences} found that shared micromobility is most effective when integrated into multimodal transport systems, particularly for first and last-mile connectivity. However, \citep{fishman2015factors} emphasised that safety concerns and regulatory barriers, such as mandatory helmet laws, negatively impact bike-share adoption. Similarly, \citep{baek2021electric} and \citep{cao2021scooter}) found that while e-scooters offer convenience and flexibility, their adoption is hindered by high pricing and perceived safety risks, suggesting that policy interventions should address both affordability and infrastructure improvements.

Studies also highlight the distinction between shared and personal micromobility usage. \citep{van2021insights} found that urban users are more likely to adopt shared micromobility, while suburban residents favour personal ownership due to lower service availability and infrastructure gaps. \citep{montes2023shared} reinforced this finding by showing that shared micromobility works best when seamlessly integrated into multimodal transport networks, particularly for short-distance urban trips. Conversely, \citep{esztergar2022assessment} found that users turn to personal micromobility when shared services are unreliable, reinforcing the need for consistent operational frameworks for shared mobility providers.

The role of micromobility in first and last-mile connectivity has also been extensively studied. \citep{van2022preferences} found that users are more likely to adopt micromobility when it significantly reduces total travel time and minimises transfer penalties associated with public transport use. \citep{montes2023shared} supported this by demonstrating that trip chaining efficiency—ensuring a seamless transition from micromobility to transit—is critical for adoption. Additionally, \citep{van2021insights} and \citep{fishman2015factors} found that bike-transit integration is influenced by external factors such as weather conditions, bike parking availability, and urban design, emphasising the need for strategic infrastructure planning.

User segmentation studies indicate that micromobility adoption varies across different demographic groups. \citep{montes2023shared,esztergar2022assessment,asgari2018stated} found that younger, urban residents are the most frequent users of shared micromobility, while older individuals and suburban residents are less likely to adopt these services due to safety concerns, unfamiliarity with digital booking platforms, and infrastructure limitations. \citep{cao2021scooter} and \citep{baek2021electric} further classified micromobility users into early adopters (tech-savvy individuals), cost-sensitive users (price-conscious travellers), and reluctant adopters (those deterred by legal or safety concerns), highlighting the need for differentiated policy approaches.

The literature also emphasises the policy and infrastructure implications for promoting micromobility adoption. \citep{montes2023shared,van2022preferences} argue that investments in bike lanes, docking stations, and transit integration are crucial for ensuring widespread adoption. \citep{fishman2015factors,esztergar2022assessment} suggest that reducing regulatory barriers, such as relaxing helmet mandates or modifying speed restrictions, could encourage higher adoption rates. Additionally, \citep{cao2021scooter} highlight the importance of equitable service distribution, ensuring that micromobility solutions reach lower-income and suburban communities, thereby reducing disparities in transportation access.

A variety of quantitative methodologies are employed to analyse micromobility adoption and mode choice behaviour. Stated preference (SP) surveys are the most commonly used method \citep{montes2023shared,van2022preferences,baek2021electric,esztergar2022assessment}, where respondents are presented with hypothetical micromobility scenarios to assess their preferences under different conditions such as travel cost, time, safety, and infrastructure availability. Discrete choice models (DCMs), including multinomial logit (MNL), mixed logit, and latent class choice models (LCCM), are widely applied to capture individual preference heterogeneity. For example, \citep{van2022preferences,esztergar2022assessment} utilise LLCMs to identify distinct groups of users based on behavioural differences, while \citep{baek2021electric,cao2021scooter} apply mixed logit models to account for variations in user sensitivity toward pricing and travel time. In addition to SP methods, some studies integrate revealed preference (RP) data, as in \citep{fishman2015factors,van2021insights}, which analyse real-world micromobility usage patterns through survey responses and transportation system data. These methodologies allow for a comprehensive understanding of micromobility adoption, distinguishing between stated user intentions and actual observed behaviours in different urban mobility contexts.

\section{ Survey design and data collection }
\subsection{ Survey design}
\begin{figure}[h]
\begin{center}
\includegraphics[width=1.0\textwidth]{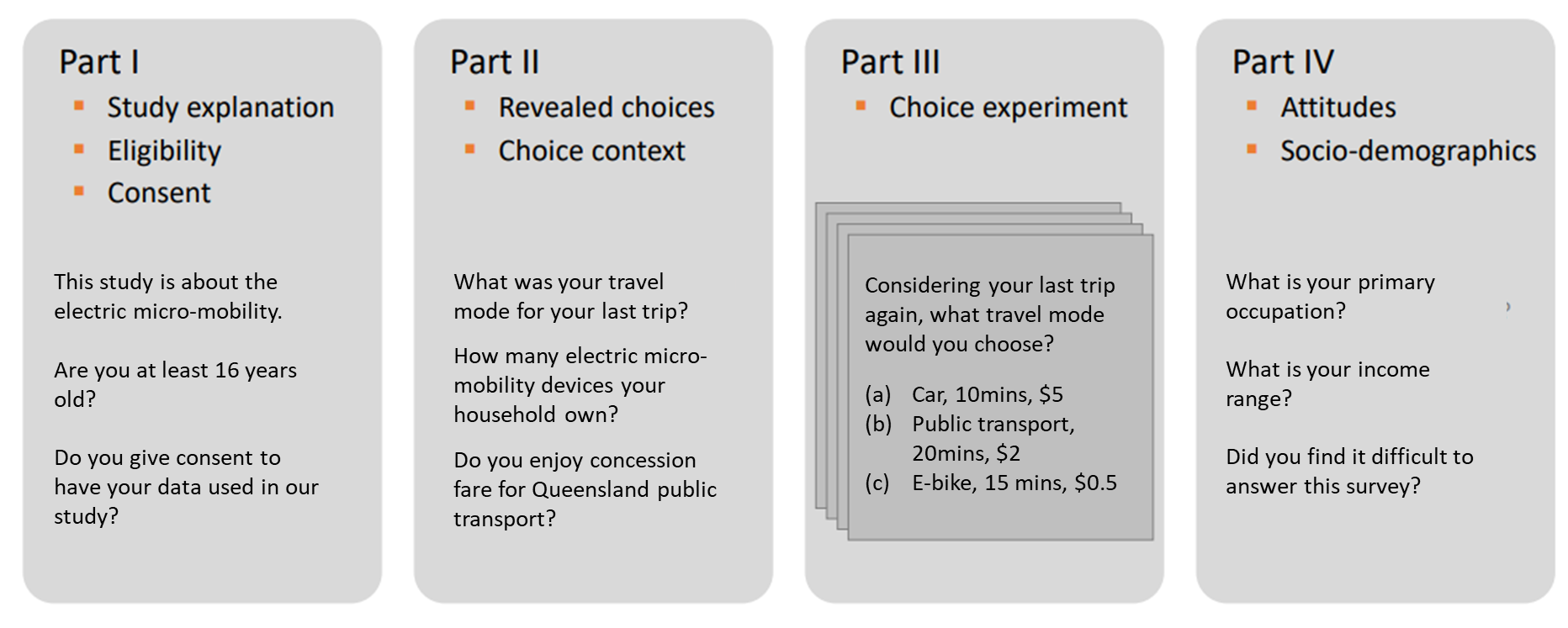}
\caption{Structure of the survey}
\label{Fig:surveyStructure}
\end{center}
\end{figure}

The questionnaire is structured into four main sections: context establishment, observation of behavioural characteristics in revealed mode choice, the design of stated preference experiments, and the collection of socio-demographic information, as illustrated in Figure~\ref{Fig:surveyStructure}
. One primary limitation of stated preference studies is their reliance on hypothetical, artificial choice scenarios. Therefore, ensuring realism in the survey design is crucial so that respondents perceive these choices as if they were making real-life decisions. Various efforts have been undertaken to enhance the survey's realism across several critical aspects, as detailed below.

\subsubsection{ Reference trip identification}
Instead of presenting respondents with a purely hypothetical trip for mode selection, the survey incorporates a reference trip as the choice context. Specifically, participants’ most recent trip, ranging from 3 to 30 kilometres, is used as the reference. Before proceeding to the stated preference experiments, they are asked to recall this trip. During the experiments, they then make the mode choices for the recalled trip, reducing cognitive load and enhancing the realism of their stated preferences.

The more recent the trip, the more realistic the choice experiments are for participants. Beyond the time dimension, the reference trip is restricted to a range of 3 to 30 kilometres. Since potential EMM journeys primarily focus on short to medium distances, the upper boundary is set to exclude unrealistically long trips for EMM use. Meanwhile, the pilot study revealed that extremely short trips could result in unrealistic attribute levels for certain alternatives and lead to an underrepresentation of public transport users, as few people choose public transport for very short trips. To address this, a lower boundary was introduced. By restricting the reference trip range, the mode choice experiments ensure more realistic and balanced scenarios for both car and public transport users.

After identifying each participant's reference trip, a screening question was introduced so that only respondents who used either a car or public transport for that trip could proceed with the survey. This filtering was not implemented in the pilot study, during which very few responses were recorded for other travel modes. Given that this study focuses on examining the potential mode shift from car and public transport users to EMM, the filtering process ensures that data collection is concentrated on the most relevant participants.

Additionally, detailed information on the reference trip is collected for subsequent mode choice experiments. In these experiments, the reference travel mode serves as the status quo option, with certain attribute levels derived directly from the actual trip details provided by respondents. Further details on this process will be discussed in a later section.

\subsubsection{ Discrete choice experiments design}
\begin{figure}[ht]
\begin{center}
\includegraphics[width=1.0\textwidth]{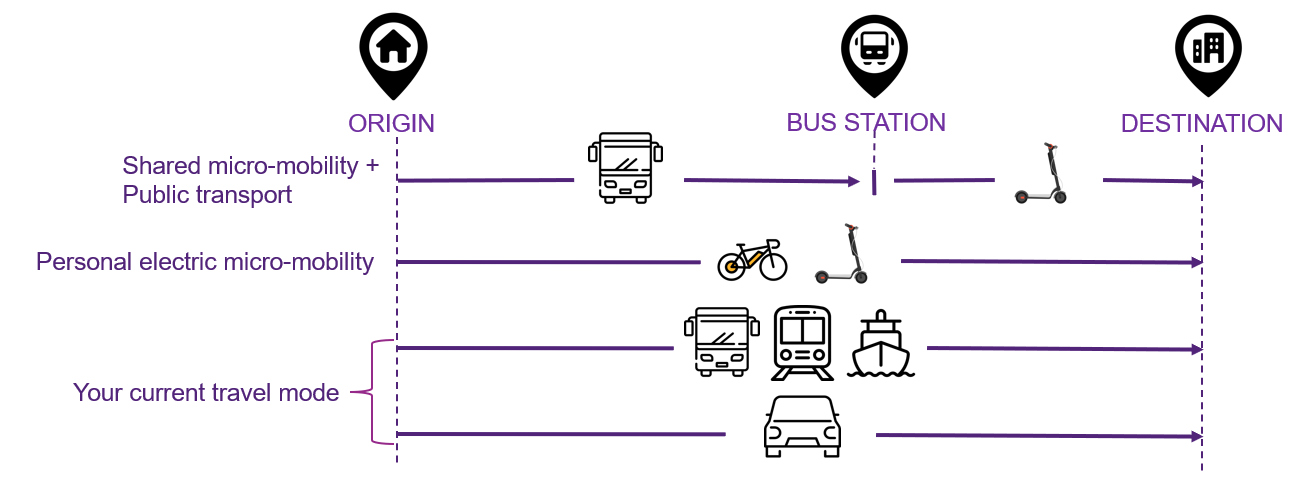}
\caption{Illustration of available options for discrete choice experiments}
\label{Fig:illustration}
\end{center}
\end{figure}

As a key component of the survey, the discrete choice experiments must be straightforward, clear, and easy to understand. Detailed instructions are provided before the stated-preference experiments to ensure that participants fully grasp the objective, which is to "carefully examine these travel modes and indicate your choices for your last trip within the range of 3 to 30 kilometres based on their attribute values." Additionally, Figure~\ref{Fig:illustration} provides an illustration that clarifies the available options of the stated choice experiments.

Participants are presented with three alternatives aligned with the study's objectives. Two of these alternatives involve EMM options: personal electric micro-mobility and shared micro-mobility integrated with public transport. We distinguish between these options because their distinct usage patterns influence travel behaviours and yield varying degrees of mode shift \citep{johnson2023impacts}. Notably, the cost of private EMM devices may significantly affect respondents’ preferences, whereas shared EMM is particularly well-suited for first- and last-mile travel as a complement to public transport. Consequently, the integrated travel mode was designed to explore the potential of multimodal trips compared to single private EMM travel. Two versions of the choice sets were developed based on respondents’ reported reference travel mode, with each set incorporating the participant's reference mode—either car or public transport. This approach facilitates a direct comparison between the proposed EMM alternatives and respondents’ actual travel behaviours, thereby enhancing the credibility of the stated preference study.

\subsubsection{Attributes and their levels}
Multiple attributes are used to describe the alternatives, with each attribute having one or more levels that may influence individuals’ mode choice behaviour. Among these, two are scenario attributes, meaning their levels remain constant across all three alternatives within a given choice scenario. The first is travel distance, which has only one level corresponding to the reported distance of the reference trip. The second is weather, which has five possible conditions. Scenario attributes are listed separately from other attributes in the choice experiments to ensure respondents recognise that they remain consistent across all alternatives.

In addition to scenario attributes, the remaining experimental attributes are categorised into four groups: travel time, travel cost, traffic congestion level, and bikeway proportion. Among these, travel time and travel cost are considered the main influential factors in mode choice experiments\citep{jaber2023preferences}. To capture their varying effects, they are further segmented into specific components, as individuals may perceive different aspects of travel time and cost differently. For instance, waiting time and in-vehicle travel time may have distinct impacts on perceived travel burden. Table~\ref{tab:attributeLevels} summarises the attribute levels for scenario attributes, which remain consistent across all alternatives, and experimental attributes, which have specific level configurations for each alternative. To better convey trip details and examine differences in respondents' perceptions, the total travel time for the public transport plus shared micromobility option is divided into two components. Specifically, waiting time is assigned to the public transport involved alternatives, representing the duration spent waiting for service upon arrival at the station. Walking time is allocated to both public transport and car trips to capture the time spent on access and egress. In addition, the proportion of bikeway indicates the extent of the riding that is completed on dedicated cycling lanes, with the remainder occurring in mixed traffic.
\begin{sidewaystable}[htbp]
\centering
\caption{Attribute levels for scenario and experimental attributes}
\label{tab:attributeLevels}
\resizebox{0.85\textwidth}{!}{%
\begin{threeparttable}
\begin{tabular}{lcccc}
\toprule   
\textbf{Attribute} & 
\makecell{\textbf{Personal e-}\\\textbf{Micro-Mobility}} &
\makecell{\textbf{Public Transport +}\\\textbf{Shared Micro-Mobility}} &
\textbf{Car} &
\makecell{\textbf{Public}\\\textbf{Transport}} \\
\midrule

\multicolumn{5}{l}{\textbf{Scenario Attributes (Common Across All Alternatives)}} \\
Weather  
& \multicolumn{4}{c}{Sunny, Windy, Cloudy, Light Rain, Heavy Rain} \\
Travel distance  
& \multicolumn{4}{c}{Reference trip distance (3--30 km)} \\[6pt]

\multicolumn{5}{l}{\textbf{Experimental Attributes (Alternative-Specific Levels)}} \\
\textit{Travel time (mins)}          
& Base value$^{\mathrm{a}}\pm25\%$ 
& Base value$^{\mathrm{a}}\pm25\%$ 
& Base value$^{\mathrm{a}}\pm25\%$ 
& Base value$^{\mathrm{a}}\pm25\%$ \\

\textit{Shared micromobility travel time (mins)}  
& n/a 
& 2, 4, 6, 8 
& n/a 
& n/a \\

\textit{Waiting time (mins)}        
& n/a 
& 1, 3, 5 
& n/a 
& \makecell[l]{\textbf{$x^{\mathrm{b}} >10$:} $x^{\mathrm{b}} \pm30\%$\\
\textbf{$x^{\mathrm{b}} <10$:} $x^{\mathrm{b}} \pm2$\\
\textbf{$x^{\mathrm{b}} =0$:} 0, 2, 4} \\

\textit{Walking time (mins)}         
& n/a 
& n/a 
& \makecell[l]{\textbf{$x^{\mathrm{b}} >10$:} $x^{\mathrm{b}} \pm30\%$\\
\textbf{$x^{\mathrm{b}} <10$:} $x^{\mathrm{b}} \pm2$\\
\textbf{$x^{\mathrm{b}} =0$:} 0, 2, 4} 
& \makecell[l]{\textbf{$x^{\mathrm{b}} >10$:} $x^{\mathrm{b}} \pm30\%$\\
\textbf{$x^{\mathrm{b}} <10$:} $x^{\mathrm{b}} \pm2$\\
\textbf{$x^{\mathrm{b}} =0$:} 0, 2, 4} \\

\textit{Running cost (AUD)} 
& $0.05 \times$ distance, $\pm25\%$ 
& \makecell[l]{\textbf{Concession:}\\
Travel distance$<9$ km: 0.5, 0.8, 1.6\\
Travel distance$\geq9$ km: 0.5, 1, 2\\
\textbf{No Concession:}\\
Travel distance$<9$ km: 0.5, 1.6, 3.2\\
Travel distance$\geq9$ km: 0.5, 2, 3.9} 
& $0.4 \times$ Travel distance, $\pm25\%$ 
& \makecell[l]{\textbf{Concession:}\\
Travel distance$<9$ km: 0.5, 0.8, 1.6\\
Travel distance$\geq9$ km: 0.5, 1, 2\\
\textbf{No Concession:}\\
Travel distance$<9$ km: 0.5, 1.6, 3.2\\
Travel distance$\geq9$ km: 0.5, 2, 3.9} \\

\textit{Parking cost (AUD)}        
& n/a 
& n/a 
& n/a 
& \makecell[l]{\textbf{$x^{\mathrm{b}} >10$:} $x^{\mathrm{b}} \pm5$\\
\textbf{$x^{\mathrm{b}} <10$:} $x^{\mathrm{b}} \pm3$\\
\textbf{$x^{\mathrm{b}} =0$:} 0, 2, 4} \\

\textit{Traffic congestion}  
& \multicolumn{4}{c}{Light, Moderate, Heavy congestion (on-road travel)} \\

\textit{Bikeway proportion}  
& \multicolumn{4}{c}{0\%, 25\%, 50\%, 75\%, 100\% (for e-Micro-Mobility routes)} \\
\bottomrule
\end{tabular}
\begin{tablenotes}
\item \textbf{Notes:} 
\item $^{\mathrm{a}}$ Travel time base value is derived from Google Maps/OpenStreetService for the reference trip distance; exact values appear in Table~\ref{tab:travelTimeBaseValues}.
\item $^{\mathrm{b}}$ Reported reference-trip value (varies by respondent).
\end{tablenotes}
\end{threeparttable}%
}
\end{sidewaystable}

Multiple real-life information sources were used to determine the attribute level values. For most travel time and travel cost attributes, instead of specifying each level explicitly, base values were first established, and then both upward and downward variations were applied to generate the full range of attribute levels. These base values were derived from a comprehensive review of the literature and simulations on the operating times and costs of various travel modes in the study area.

Specifically, for in-vehicle travel time, six base values corresponding to six trip distance segments (ranging from 5 km to 30 km in 5 km increments) were proposed for each travel mode. A 15 km radius centred on Brisbane was defined, and bus stops within this area were identified using the General Transit Feed Specification (GTFS) dataset available through Queensland’s open data portal, yielding 5,670 bus stops. Trip simulations between two randomly selected bus stops were then conducted using the Google Maps API across various travel modes—including driving, bicycling, and transit. After 2,000 simulations, the average travel time for each mode at each distance segment was calculated to establish the base values (See Table~\ref{tab:travelTimeBaseValues}).
\begin{table}[htbp]
\centering
\caption{Base value of travel times (mins) by distance segment and alternative}
\label{tab:travelTimeBaseValues}
\begin{threeparttable}
\begin{tabular}{lcccc}
\toprule
\textbf{Distance (km)} & 
\makecell{\textbf{Personal e-}\\\textbf{Micro-Mobility}} &
\makecell{\textbf{Public Transport +}\\\textbf{Shared Micro-Mobility}} &
\textbf{Car} &
\makecell{\textbf{Public}\\\textbf{Transport}} \\
\midrule
$<$ 5   & 10.34 &  7.98 &  6.42 & 18.44 \\
5--10   & 23.25 & 26.71 & 12.95 & 36.33 \\
10--15  & 37.06 & 32.59 & 18.44 & 50.12 \\
15--20  & 51.35 & 45.04 & 22.41 & 66.23 \\
20--25  & 65.68 & 49.80 & 25.64 & 72.07 \\
25--30  & 79.66 & 61.31 & 29.08 & 86.91 \\
\bottomrule
\end{tabular}
\begin{tablenotes}
\item \textbf{Note:} All travel times are given in minutes. The base values were calculated as the average from 2,000 simulations.
\end{tablenotes}
\end{threeparttable}
\end{table}

While base values of travel times for cars and public transport were directly derived from Google Maps API, additional calculations were required for the EMM-specific alternatives, as Google Maps does not include an EMM mode. To address this, an API from OpenStreetService—which supports e-bikes, a common form of EMM—was utilised. Trips were simulated in a manner similar to the Google Maps API, and the travel times for bicycles and e-bikes were compared. The analysis showed that, across all distance segments, e-bikes took approximately 0.8 times the travel time of regular bicycles. Based on this ratio, the base value for personal electric micro-mobility was set to 0.8 times the average bicycle travel time extracted from Google Maps API.

The total travel time for the multimodal alternative is defined as the average public transport travel time minus the walking time required for access and egress, which can be obtained from the Google Maps API. Simulation results indicate that the egress walking segment typically takes around 10 minutes, whereas covering the same distance using EMM would take approximately 2 minutes. Based on this, four levels for shared micro-mobility travel time were established: 2, 4, 6, and 8 minutes. Accordingly, the public transport travel time component is assigned three levels, representing low, moderate, and high micro-mobility usage, calculated as the base total travel time for the multimodal alternative minus 2, 5, and 8 minutes, respectively.

A comprehensive review of the local travel market was conducted to establish base travel cost values for each alternative, incorporating data from government and insurance reports. Manual adjustments were also applied to enhance the realism of the choice experiments. Specifically, for public transport, Brisbane introduced a 50-cent fare scheme trial during the study period, standardising the cost of all public transport trips at 50 cents. Accordingly, the lowest public transport fare level was set to 50 cents. Additionally, parking costs for the car alternative were determined based on participants' reported expenses. Furthermore, traffic congestion levels and bikeway proportions were assigned three and five levels, respectively, to assess their impact on mode choice behaviour.

\subsubsection{Efficient experimental design}
In each discrete choice experiment, a specific configuration of attribute levels is presented, referred to as a choice situation. A full factorial design, which includes all possible attribute level combinations, generates the maximum number of choice situations without repetition. Mathematically, a full factorial design produces $\prod_{j=1}^{J} \prod_{k=1}^{K_j} \ell^{jk}$ choice situations, where $J$ represents the number of alternatives, each with $K_j$ attributes, and each attribute $k \in K_j$ has $l_jk$ levels. 

In our case, a full factorial design would result in millions of choice situations for both car and public transport users, making it impractical and leading to unrealistic scenarios. To address this, we applied a D-efficient design \citep{mcfadden1974measurement} to optimise observable trade-offs (minimising the sample size while the t-ratios are still statistically significant \citep{rose2009constructing,de2024modelling}) critical for choice modelling while maintaining the realism of choice scenarios.

To design an efficient experiment, it's essential to have preliminary information on factor coefficients. The more closely these prior estimates align with the actual parameters, the more effective the design becomes. However, exact coefficients are only determined through choice modelling after data collection. As these coefficients are unknown during the design stage, the prior information must serve as the best estimates of the real parameters. These priors can be sourced from existing literature on similar projects or from data collected in a pilot study. In this study, the latter strategy was implemented by conducting a pilot study first to derive more precise priors close to the actual coefficients. For the pilot study itself, though, uninformative priors were utilised, as no field estimates were available. These uninformative priors are expressed as small values, with either a positive or negative sign to reflect the anticipated influence of each factor on the respective alternative (e.g., -0.001 to indicate the negative effect of travel cost on car selection). 

For this study, Ngene \citep{choicemetrics2012ngene} was used to generate an efficient design. The software iteratively builds multinomial logit (MNL) models based on different designs, calculates the D-error for each, and retains the design with the lowest D-error. In the pilot study, data were collected from 145 respondents, and MNL models were estimated to extract informative priors. These priors were then used to refine the efficient design, which was subsequently applied in the main study. Notably, instead of using a single point estimate for the informative priors, a Bayesian distribution was employed, incorporating both the coefficient and its standard deviation. This approach accounts for uncertainty and enhances the reliability of the experimental design. Due to the complexity of this study, the experimental design remains large even after efficient design optimisation, consisting of 60 choice situations. To manage cognitive burden and ensure robust data collection, a blocking strategy was applied, dividing the design into 10 blocks. Each respondent was randomly assigned to one block, resulting in six choice tasks per participant.

\subsection{Data collection}
The survey was conducted in Brisbane through three rounds of online distribution—the pilot, main, and resampling rounds—between June and October 2024. The first page of the survey included an introduction to the study, followed by a consent question to confirm participants’ willingness to proceed. A participant information sheet was also provided, detailing the study’s purpose, data storage practices, and assurances of anonymity. Participants were informed of their right to withdraw at any time. SurveyEngine, a professional online panel provider, was responsible for recruiting participants and collecting data.

A screening process was implemented to ensure the recruitment of a relevant sample. In accordance with Brisbane government regulations, the minimum age for riding micro-mobility devices without supervision is 16 years, which was set as the eligibility threshold for survey participation. Additionally, respondents were asked early in the survey to recall their reference trip and identify their primary travel mode. Since this study focuses exclusively on car and public transport users, individuals who reported using other modes were screened out and did not proceed further in the survey.

To ensure representative sampling from the study area, age and gender quotas were applied to align with the local demographic distribution. Additionally, two attention check questions were incorporated into the survey. The first required respondents to select a specific response, ensuring they were paying attention. The second involved a simple warm-up choice before the main experiment. Only participants who passed the first check were included in the study, while resampling was conducted for those who failed the second attention check.

The study collected a total of 1,671 responses across three rounds. During the pilot study, a small number of participants using travel modes other than car and public transport were recorded. After screening out these respondents, along with other invalid samples as described earlier, 1,643 valid responses remained. These were categorised into two groups: 1,225 car users and 418 public transport users. Each respondent completed six stated preference choice scenarios, resulting in 7,350 observations for car users and 2,508 observations for public transport users, which were subsequently used to develop the choice models.

\subsubsection{Sample profile}
Socio-demographic information was collected in the final section of the questionnaire, following the argument that presenting this section before the stated preference experiments could heighten respondents’ self-awareness and potentially influence their choices in an unnatural manner. Table~\ref{tab:sampleProfile} presents the detailed sample profiles, including demographic characteristics and reference trip attributes. Additionally, the marginal distributions of gender and age are reported for both the sample and the local population, with the latter sourced from \citep{ABS2021}.

\begin{table}[htbp]
\centering
\caption{Sample profile and local population distribution}
\label{tab:sampleProfile}
\resizebox{0.85\textwidth}{!}{%
\begin{tabular}{lccc}
\toprule
\textbf{Category} & \textbf{Frequency} & \textbf{Percentage (\%)} & \textbf{Local Distribution (\%)} \\
\midrule
\multicolumn{4}{l}{\textbf{Gender}} \\
Male                   & 799  & 48.6 & 49.2 \\
Female                 & 836  & 50.9 & 50.8 \\
Other                  & 5    & 0.3  & --   \\
Prefer not to say      & 3    & 0.2  & --   \\
\midrule
\multicolumn{4}{l}{\textbf{Age}} \\
16--19 years          & 150  & 9.1  & 8.4  \\
20--29 years          & 294  & 17.9 & 15.4 \\
30--39 years          & 307  & 18.7 & 16.2 \\
40--49 years          & 254  & 15.5 & 15.4 \\
50--59 years          & 250  & 15.2 & 15.0 \\
60--69 years          & 229  & 13.9 & 13.3 \\
70--79 years          & 130  & 7.9  & 10.3 \\
80 years and older     & 29   & 1.8  & 6.4  \\
\midrule
\multicolumn{4}{l}{\textbf{Employment}} \\
Work full-time         & 807  & 49.1 & 55.8 \\
Work part-time         & 255  & 15.5 & 30.5 \\
Unemployed             & 73   & 4.4  & 5.4  \\
Self-employed          & 98   & 6.0  & --   \\
Student                & 113  & 6.9  & --   \\
Retired                & 235  & 14.3 & --   \\
Homemaker              & 42   & 2.6  & --   \\
Other                  & 20   & 1.2  & 8.3  \\
\midrule
\multicolumn{4}{l}{\textbf{Weekly Income}} \\
Less than \$500        & 304  & 18.5 & 31.1 \\
\$500--\$999           & 525  & 32.0 & 23.6 \\
\$1,000--\$1,999       & 571  & 34.8 & 26.9 \\
More than \$2,000      & 243  & 14.8 & 11.7 \\
\midrule
\multicolumn{4}{l}{\textbf{Education}} \\
Bachelor Degree level and above & 893  & 54.4 & 21.9 \\
\bottomrule
\end{tabular}
}
\begin{tablenotes}
\item \textbf{Note:} Data of local distribution comes from Australian Bureau of Statistics Census 2021, of Queensland.
\end{tablenotes}
\end{table}

For weekly income and education level, the sample in this study shows higher proportions than the local population distribution, reflecting a reasonable technical bias due to the online survey approach. The remaining data in Table~\ref{tab:sampleProfile} indicates that the sample is broadly representative of the local population. Among all respondents, 90.9\% reported holding a valid driver’s license, and 91.8\% had access to a car, with 43.1\% owning more than one vehicle. Regarding EMM, 55.6\% of respondents had experience using EMM, and 36.1\% reported owning EMM devices, with 28.3\% of these households owning more than one. Additionally, 31.9\% of respondents were members of one of Brisbane’s two main shared micromobility service providers during the survey period. The reference trip distances ranged from 3 to 30 kilometres, with an average of 13.5 kilometres and a median of 12 kilometres. Among all trips, 36.5\% were personal, 30\% employment-related, 23.1\% recreational, and 6\% education-related. Cars were used for 73\% of the 1,671 journeys (1,225 trips), while 307 were public transport trips.

\section{ Model framework}

The Latent Class Choice Model (LCCM) extends the traditional Multinomial Logit (MNL) model by accounting for individual heterogeneity through a finite number of latent classes. First introduced by \citep{lazarsfeld1968latent} and further developed by \citep{greene2003latent}, LCCM captures both observable choice-related attributes and latent heterogeneity, which is not directly observable from choice behaviour, such as variations in respondents’ socio-demographic characteristics. The model consists of two key components: the class membership model, which estimates the probability of an individual belonging to each latent class based on latent factors (e.g., socio-demographic characteristics in this study), and the class-specific choice model, which estimates the probability of selecting each alternative within a given choice scenario for a specific class.In essence, the two-level framework is built by nesting two MNL models: one to estimate the probability of class assignment, and the other to estimate the choice probability within each class. Model estimation is carried out using the \texttt{apollo} package in R \citep{hess2019apollo}, which facilitates an iterative estimation process whereby the posterior probabilities from the class membership model and the class-specific choice models are updated in turn until convergence.

\paragraph{Class-Specific Choice Model.}  
Within each latent class \( s \) (\( s = 1,2,\dots,S \)), the utility for individual \( n \) choosing alternative \( i \) in choice situation \( t \) is given by
\begin{equation}
U_{nti|s} = \beta_s' X_{nti} + \epsilon_{nti|s},
\end{equation}
where \( X_{nti} \) is a vector of observed attributes of the alternatives (i.e., travel time, travel cost), \( \beta_s \) is a vector of class-specific parameters, and \( \epsilon_{nti|s} \) is an error term assumed to follow an Extreme Value Type I (Gumbel) distribution. Under the standard MNL assumption within classes, the probability that individual \( n \) in class \( s \) chooses alternative \( i \) in scenario \( t \) is:
\begin{equation}
P(y_{nti}=i \mid s) = \frac{\exp(\beta_s' X_{nti})}{\sum_{j \in C_t} \exp(\beta_s' X_{ntj})},
\end{equation}
where \( C_t \) denotes the set of available alternatives in situation \( t \) (i.e., personal electric micro-mobility).

\paragraph{Class Membership Model.}  
The class membership model assigns individuals to latent classes based on specific individual covariates that can reflect heterogeneity among latent classes \( Z_n \) (i.e., age, gender). The probability that individual \( n \) belongs to class \( s \) is modelled as:
\begin{equation}
M_n(s) = \frac{\exp(\gamma_s' Z_n)}{\sum_{s'=1}^{S} \exp(\gamma_{s'}' Z_n)},
\end{equation}
with \( \gamma_s \) representing the parameters associated with class \( s \).

\paragraph{Overall Model Probability.}  
The overall probability of observing an individual’s choice is obtained by integrating over the latent classes:
\begin{equation}
P(y_{nti}=i) = \sum_{s=1}^{S} M_n(s) \, P(y_{nti}=i \mid s).
\end{equation}
This formulation accounts for both heterogeneity in preferences through segmentation into latent classes and heterogeneity in choice behaviour through the class-specific choice models.

\paragraph{Iterative Estimation Procedure.}
The modelling approach involves an iterative estimation of the LCCM in which the posterior probabilities from the class membership model and the class-specific choice models are mutually updated. The procedure proceeds as follows:

\textbf{Step 1:} Initialise the parameters \( \{\beta_s, \gamma_s\} \) for each latent class. In the implementation using the \texttt{apollo} package in R, the initial values for the class-specific parameters \( \beta_s \) and the membership model parameters \( \gamma_s \) are typically set to zero, or to small values derived from previous models. As model complexity increases (e.g., with the introduction of additional latent classes), the initial values are often based on outcomes from simpler models (such as an MNL model) to guide the iterative estimation procedure.

\textbf{Step 2:} With the current estimates, compute the posterior probability that individual \( n \) belongs to class \( s \) using Bayes’ rule:
\begin{equation}
\pi_{ns} = \frac{M_n(s) \prod_{t} P(y_{nti} \mid s)}{\sum_{s'=1}^{S} M_n(s') \prod_{t} P(y_{nti} \mid s')}.
\end{equation}

\textbf{Step 3:} Re-estimate the class-specific choice model parameters \( \beta_s \) by maximising the weighted likelihood, where each individual’s contribution is weighted by the posterior probability \( \pi_{ns} \).

\textbf{Step 4:} Update the class membership model parameters \( \gamma_s \) by re-estimating the membership model with the updated \( \pi_{ns} \) as the dependent variable.

\textbf{Step 5:} Repeat Steps 2–4 until convergence is reached, as indicated by negligible changes in the log-likelihood or parameter estimates between iterations.

The final parameter estimates obtained through this iterative process constitute the optimal configuration of the LCCM. This method accounts for the interdependence between the class membership and the class-specific choice components, resulting in a robust segmentation of the population and reliable estimates of the latent class-specific choice behaviour.

\section{ Results}


\subsection{Identification of latent heterogeneity}
Initially, MNL models were estimated separately for the car user experiments and the public transport user experiments, serving as the baseline models, or equivalently, one-class LCCMs. Although these MNL models do not account for heterogeneity across the sample—resulting in less precise estimates—they still provide valuable insights into overall choice behaviour. For instance, while the absolute magnitudes of the factors' impacts may be imprecise, the sign of each coefficient reliably indicates whether a factor exerts a positive or negative influence on the choice of a particular alternative over the whole sample. Several versions of the MNL model were developed, ranging from a basic specification that included only alternative-specific constants (ASCs) and experimental attributes to more comprehensive models that also incorporated individual-specific characteristics.

Specifically, the individual characteristic factors that proved significant in the MNL models are considered the source of heterogeneity among latent classes. Initially, all individual characteristic factors collected from the survey were input into the choice analysis function provided by \texttt{apollo} to assess their impact on choice behaviour. Subsequently, statistically significant factors influencing choice behaviour were identified separately for car users and public transport users. These significant factors, together with the class-specific constants, were then incorporated into the LCCMs to drive the class membership model that assigns individuals to latent segments. Table~\ref{tab:classMembershipCovariates} presents the detailed covariates used for estimating the class membership model for both car users and public transport users.
\begin{table}[h!]
\centering
\caption{Covariates used for estimating the class membership model for both car users and public transport users.}
\label{tab:classMembershipCovariates}
\begin{tabular}{ll}
\toprule
\textbf{LCCMs} & \textbf{Covariates conditions} \\
\midrule
Car users & Micromobility experience \\
          & Micromobility ownership \\
          & Female \\
          & Age over 50 \\
          & Public transport concession \\
\midrule
Public transport users & Micromobility experience \\
                        & Micromobility ownership \\
                        & Shared micromobility membership \\
                        & Female \\
                        & Age between 20 to 49 \\
                        & Public transport concession \\
                        & Full-time work \\
                        & Weekly income less than \$1,000 \\
\bottomrule
\end{tabular}
\end{table}

\subsection{Number of latent classes for LCCMs}
Departing from the MNL models, LCCMs with an increasing number of latent classes were gradually built and tested. Determining the optimal number of latent classes is crucial and should first consider the interpretability of the model outputs. To support this decision, common performance indicators such as the Akaike Information Criterion (AIC), the Bayesian Information Criterion (BIC), and goodness-of-fit measures are used \citep{louviere2000stated,shen2014latent,zhou2020analysing}. In particular, AIC and BIC assess whether the improvement in model fit justifies the additional parameters, thereby mitigating potential overfitting. The goodness-of-fit measure employed in this study is the Rho-squared relative to an equal shares model, denoted as \( \rho^2_{\text{ES}} \) and defined by  

\begin{equation}
\rho^2_{\text{ES}} = 1 - \frac{L_{\text{final}}}{L_{\text{ES}}},
\end{equation}

where \( L_{\text{final}} \) is the log-likelihood of the estimated model with all explanatory variables, and \( L_{\text{ES}} \) is the log-likelihood of the equal shares model, in which each alternative is assumed to have an equal probability of being chosen.

Additionally, the relative sizes of the latent segments can provide important guidance in determining the optimal number of classes in the LCCM. If a latent class contains only a small fraction of respondents, the limited data available for that segment may lead to unreliable estimation of the class-specific MNL model or even estimation failures. Literature suggests that a latent class comprising less than 15\% of the sample in a two-class model, or less than 10\% in models with three or more classes, may be indicative of a poor model fit \citep{sinha2021practitioner}. Table~\ref{tab:modelPerformanceCriteria} summarises the evaluation metrics based on these criteria for both the car user LCCMs and the public transport user LCCMs. It is important to note that the LCCM for both three-class car users and two-class public transport users, as shown in this table, represents the final version. Attributes that were consistently found to be insignificant across all classes have been excluded, leading to fewer estimated parameters in the final version.
\begin{table}[ht]
\centering
\caption{Model Performance for Number of Segments}
\label{tab:modelPerformanceCriteria}
\resizebox{0.85\textwidth}{!}{%
\begin{threeparttable}
\textbf{Car Users LCCMs}
\vspace{0.3cm}

\begin{tabular}{lcccc}
\toprule
\multirow{2}{*}{\textbf{Criteria}} & \multicolumn{4}{c}{\textbf{Number of Segments}} \\
\cmidrule(lr){2-5}
        & \textbf{1 (MNL)} & \textbf{2} & \textbf{3} & \textbf{4} \\
\midrule
Number of parameters & 34 & 58 & 63 & 122 \\
Log likelihood & -6315.34 & -5803.04 & -5613.82 & -5461.72 \\
\( \rho^2_{\text{ES}} \) &  0.1353 & 0.2813 & 0.3048 & 0.3236 \\
AIC & 12698.67 & 11722.09 & 11353.63 & 11167.45 \\
BIC & 12929.94 & 12122.43 & 11788.49 & 12009.55 \\
Number of individuals & \multicolumn{4}{c}{1225} \\
Number of observations & \multicolumn{4}{c}{7350} \\
Segment size & -- & \begin{tabular}[c]{@{}c@{}}Segment 1: 45.96\% \\ Segment 2: 54.04\%\end{tabular} & \begin{tabular}[c]{@{}c@{}}Segment 1: 36.70\% \\ Segment 2: 41.68\% \\ Segment 3: 21.62\%\end{tabular} & \begin{tabular}[c]{@{}c@{}}Segment 1: 26.98\% \\ Segment 2: 36.85\% \\ Segment 3: 10.77\% \\ Segment 4: 25.39\%\end{tabular} \\
\bottomrule
\end{tabular}

\vspace{0.5cm}
\textbf{Public Transport Users LCCMs}
\vspace{0.3cm}

\begin{tabular}{lcccc}
\toprule
\multirow{2}{*}{\textbf{Criteria}} & \multicolumn{4}{c}{\textbf{Number of Segments}} \\
\cmidrule(lr){2-5}
        & \textbf{1 (MNL)} & \textbf{2} & \textbf{3} & \textbf{4} \\
\midrule
Number of parameters & 34 & 25 & 93 & 127 \\
Log likelihood & -2383.27 & -2265.71 & -2123.05 & -2047.19 \\
\( \rho^2_{\text{ES}} \) &  0.0435 & 0.1777 & 0.2295 &  0.257 \\
AIC & 4834.53 & 4581.41 & 4432.11 & 4348.38 \\
BIC & 5029.24 & 4727.09 & 4974.04 & 5088.44 \\
Number of individuals & \multicolumn{4}{c}{418} \\
Number of observations & \multicolumn{4}{c}{2508} \\
Segment size & -- & \begin{tabular}[c]{@{}c@{}}Segment 1: 77.92\% \\ Segment 2: 22.08\%\end{tabular} & \begin{tabular}[c]{@{}c@{}}Segment 1: 19.98\% \\ Segment 2: 56.57\% \\ Segment 3: 23.46\%\end{tabular} & \begin{tabular}[c]{@{}c@{}}Segment 1: 21.58\% \\ Segment 2: 13.09\% \\ Segment 3: 18.77\% \\ Segment 4: 46.55\%\end{tabular} \\
\bottomrule
\end{tabular}
\end{threeparttable}%
}
\end{table}

For car users, the modelling results show that \( \rho^2_{\text{ES}} \) increases as the number of latent segments increases. However, this improvement does not necessarily imply that a 4-class model is optimal. In contrast to AIC, which often selects models with better fitness, BIC applies a tougher penalty for additional model complexity. Moreover, the 4-class model includes a residential segment comprising only about 10\% of the sample, which raises concerns regarding the reliability of parameter estimates due to the disproportionate number of parameters relative to the limited data in that segment. Bootstrap testing further confirmed these concerns by revealing unstable estimates for the 4-class model. Consequently, the 3-class model is preferred, as it provides a better balance between model fit and parsimony for the car users' choice model.

Public transport users' choice models exhibit a similar trend when testing different numbers of latent classes. The goodness-of-fit improves as more classes and parameters are introduced. However, when the number of classes increases to four, a residential segment comprising less than 10\% of the sample emerges, resulting in unstable model estimation. While the transition from two to three classes improves AIC and goodness-of-fit measures, the BIC value significantly increases, indicating potential overfitting. Furthermore, the three-class model offers limited interpretative benefits compared to the two-class model. Although it subdivides the larger segment identified in the two-class model, the resulting subgroups do not exhibit significant differences, as reflected in the negligible differences in class intercepts. Based on both statistical indicators and interpretative considerations, the two-class LCCM model was selected for further analysis of public transport users’ choices.

\subsection{Parameters estimated}
Table~\ref{tab:carLCCM} and Table~\ref{tab:PTLCCM} report the parameter estimates of the selected LCCM for car users' mode choice behaviour and public transport users' mode choice behaviour, respectively. Only parameters that were found to be significant within a 95\% confidence interval for at least one segment are included in the final reported LCCMs. For both groups, the parameters can be categorised into two main sets. One set consists of attributes that are directly related to the trip characteristics, such as travel cost and travel time. The other set includes additional attributes—such as scenario attributes (weather conditions, and travel distance) and alternative-specific constants (ASCs)—that capture inherent preferences. Notably, the latter group tends to have a greater impact on mode choice across all segments.

\begin{sidewaystable}[htbp]
\centering
\rotatebox{180}{%
\begin{minipage}{\textheight}
\centering
\caption{Parameter estimates for car users three-class LCCM}
\label{tab:carLCCM}
\begin{threeparttable}
\begin{tabular}{lrr  rr  rr}
\toprule
 & \multicolumn{2}{c}{\makecell{\textbf{Segment 1} \\ (36.70\%) \\ Multimodal trip supporters}} 
 & \multicolumn{2}{c}{\makecell{\textbf{Segment 2} \\ (41.68\%) \\ Micromobility resistant}} 
 & \multicolumn{2}{c}{\makecell{\textbf{Segment 3} \\ (21.62\%) \\ Personal micromobility lovers}} \\
\cmidrule(lr){2-3}\cmidrule(lr){4-5}\cmidrule(lr){6-7}
\textbf{Parameter} & \textbf{Coef.} & \textbf{Rob. t.rat.} 
 & \textbf{Coef.} & \textbf{Rob. t.rat.} 
 & \textbf{Coef.} & \textbf{Rob. t.rat.} \\
\midrule
asc\_pmm\tnote{a}      & 0.22 & 0.33 & \textbf{-2.97} & \textbf{-4.67} & 0.44 & 0.59 \\
asc\_ptsmm\tnote{b}    & 0.72 & 1.38 & \textbf{-3.73} & \textbf{-4.03} & -1.34 & -1.59 \\
asc\_car      & 0.00 & --   & 0.00   & --    & 0.00 & --    \\
b\_travelDistance\_pmm   & -0.05 & -1.55 & \textbf{-0.12} & \textbf{-2.32} & 0.01 & 0.08 \\
b\_travelDistance\_ptsmm & 0.00  & 0.04  & \textbf{-0.25} & \textbf{-2.33} & 0.07 & 0.45 \\
b\_windy\_pmm   & -0.31 & -1.79 & \textbf{-1.47} & \textbf{-3.03} & -0.54 & -1.63 \\
b\_cloudy\_pmm   & -0.14 & -0.77 & -0.59  & -1.87 & 0.04  & 0.11 \\
b\_Rainy\_pmm  & \textbf{-0.83} & \textbf{-3.77} & \textbf{-1.45} & \textbf{-3.61} & \textbf{-1.92} & \textbf{-3.86} \\
b\_Rainy\_ptsmm & \textbf{-0.54} & \textbf{-3.38} & \textbf{-1.24} & \textbf{-2.50} & \textbf{-1.54} & \textbf{-2.76} \\
\\[4pt]
b\_inVehicleTravelTime\_ptsmm & \textbf{-0.03} & \textbf{-2.68} & 0.00  & -0.04 & -0.01 & -0.25 \\
b\_inVehicleTravelTime\_car   & \textbf{-0.03} & \textbf{-2.95} & \textbf{-0.07} & \textbf{-3.11} & -0.04 & -0.58 \\
b\_micromobilityTravelTime\_pmm   & \textbf{-0.02} & \textbf{-2.81} & \textbf{-0.06} & \textbf{-3.11} & -0.01 & -0.71 \\
b\_waitTime\_ptsmm   & \textbf{-0.05} & \textbf{-2.00} & 0.03  & 0.33  & -0.03 & -0.64 \\
b\_walkTime\_car   & \textbf{-0.02} & \textbf{-2.03} & -0.02 & -1.85 & -0.04 & -0.89 \\
b\_runningCost\_ptsmm   & -0.08 & -1.27 & -0.22 & -1.72 & \textbf{-0.45} & \textbf{-2.62} \\
b\_runningCost\_car     & \textbf{-0.08} & \textbf{-2.00} & \textbf{-0.31} & \textbf{-3.36} & 0.08  & 0.39 \\
b\_pc\_car     & -0.01 & -1.16 & \textbf{-0.05} & \textbf{-9.41} & -0.21 & -1.37 \\
b\_cl\_car     & 0.11  & 0.36  & -0.35 & -0.62 & -1.15 & -1.73 \\
\bottomrule
\end{tabular}
\begin{tablenotes}
\footnotesize
\item[a] pmm stands for alternative "personal e-micromobility".
\item[b] ptsmm stands for alternative "public transport with shared micromobility".
\item Note: Coefficients are bolded when their robust t-ratios are significant at the 0.05 level (i.e., $|t| \geq 1.96$).
\end{tablenotes}
\end{threeparttable}
\end{minipage}%
}
\end{sidewaystable}

\begin{table}[htbp]
\centering
\caption{Parameter estimates for public transport users two-class LCCM}
\label{tab:PTLCCM}
\begin{threeparttable}
\begin{tabular}{lrr  rr}
\toprule
 & \multicolumn{2}{c}{\makecell{\textbf{Segment 1}\\(77.92\%)\\Micromobility Adopter}} 
 & \multicolumn{2}{c}{\makecell{\textbf{Segment 2}\\(22.08\%)\\Resistant}} \\
\cmidrule(lr){2-3}\cmidrule(lr){4-5}
\textbf{Parameter} & \textbf{Coef.} & \textbf{Rob. t.rat.} 
 & \textbf{Coef.} & \textbf{Rob. t.rat.} \\
\midrule
asc\_pmm\tnote{a}      & \textbf{1.64}  & \textbf{7.56}  & \textbf{-3.52} & \textbf{-4.67} \\
asc\_ptsmm\tnote{b}    & \textbf{1.50}  & \textbf{8.02}  & \textbf{-2.12} & \textbf{-3.41} \\
asc\_pt\tnote{c}       & 0.00           & --             & 0.00           & -- \\
b\_travelDistance\_pmm   & \textbf{0.04}  & \textbf{2.44}  & 0.32           & 1.71 \\
b\_Rainy\_pmm  & \textbf{-1.40} & \textbf{-8.96} & -1.34          & -1.20 \\
b\_Rainy\_ptsmm & \textbf{-1.19} & \textbf{-7.74} & -0.54          & -0.89 \\
\\[4pt]
b\_micromobilityTravelTime\_pmm  & \textbf{-0.02} & \textbf{-4.25} & -0.15          & -1.73 \\
b\_runningCost\_ptsmm  & \textbf{-0.15} & \textbf{-2.20} & -0.53          & -1.11 \\
b\_runningCost\_pt     & \textbf{-0.19} & \textbf{-2.01} & -0.40          & -1.48 \\
\bottomrule
\end{tabular}
\begin{tablenotes}
\footnotesize
\item[a] pmm stands for alternative "personal e-micromobility".
\item[b] ptsmm stands for alternative "public transport with shared micromobility".
\item[c] pt stands for alternative "public transport".
\item Note: Coefficients are bolded when their robust t-ratios are significant at the 0.05 level (i.e., $|t| \geq 1.96$).
\end{tablenotes}
\end{threeparttable}
\end{table}

In particular, for segment 2 of public transport users, most of the attribute estimates are statistically insignificant, whereas the ASCs exhibit a strong negative effect on the likelihood of choosing the proposed micromobility alternatives. This suggests that individuals in this segment display considerable resistance to changing their travel mode, with their decisions being less influenced by external factors and more driven by a personal reluctance to adopt micromobility. However, the size of this segment only counts for around one fifth of the whole public transport users group, with another group of public transport users hold great potential of accepting micromobility usage.

Furthermore, the running cost of personal micromobility is found to be insignificant across all segments for both car and public transport users, likely because this cost is negligible compared to other travel cost factors. In contrast to expectations, the bikeway attribute does not show a significant impact under any scenario. This lack of significance may be attributed to limited riding experience and the restricted coverage of bikeways in the current city, which diminishes the perceived importance of this factor.

In general, a recurring trend can be observed. Rainy conditions tend to adversely affect most options, while both the cost and duration of travel typically show the expected negative results. Although the parameter estimates show consistent signs across different segments, their responsiveness to various features varies. This variation reflects heterogeneity in how each segment perceives and responds to these factors, ultimately influencing their mode choice behaviour. The magnitude of parameters indicates the extent of their impact; however, the absolute values are not meaningful by themselves, as they will be analyzed as elasticities in a subsequent section.

\subsection{Model prediction}
The predicted choice probabilities were generated using Apollo’s built-in prediction function, which applies the parameters estimated for the choice models of each specific latent class. The resulting table (Table~\ref{tab:prediction}) presents, for each class, the proportion of respondents predicted to choose each alternative. These predicted shares offer clear insights into the heterogeneity of mode choice preferences across segments for both car users and public transport users. The behaviour-oriented predictions are then used to label each class, ensuring a clear and direct representation of each segment's mode choice preference.
\begin{table}[htbp]
\centering
\caption{Predicted Choice Probabilities by Segment}
\label{tab:prediction}
\resizebox{0.85\textwidth}{!}{%
\begin{threeparttable}
\begin{tabular}{lccc}
\toprule
\multicolumn{4}{c}{\textbf{Car Users}} \\
\midrule
\textbf{Prediction} & \makecell{\textbf{Segment 1}\\ Multimodal trip supporter\\ (36.70\%)} 
                    & \makecell{\textbf{Segment 2}\\ Micromobility resistant\\ (41.68\%)} 
                    & \makecell{\textbf{Segment 3}\\ Personal micromobility lover\\ (21.62\%)} \\
\midrule
personal micromobility    & 24\% & 5\%  & 67\% \\
\text{Public Transport +}\\\text{Shared Micro-Mobility}  & 47\% & 4\%  & 17\% \\
car    & 29\% & 90\% & 16\% \\
\midrule
\multicolumn{4}{c}{\textbf{Public Transport Users}} \\
\midrule
\textbf{Prediction} & \makecell{\textbf{Segment 1}\\ Micromobility adopter\\ (77.92\%)} 
                    & \makecell{\textbf{Segment 2}\\ Micromobility resistant\\ (22.08\%)} 
                    & \\
\midrule
personal micromobility    & 42\% & 2\%  & \\
\text{Public Transport +}\\\text{Shared Micro-Mobility}  & 42\% & 8\%  & \\
public transport     & 16\% & 89\% & \\
\bottomrule
\end{tabular}
\end{threeparttable}%
}
\end{table}

The predicted choice probabilities highlight the significant potential for mode shift towards EMM usage among both car and public transport users. However, the analysis also identifies a segment within each group that remains resistant to EMM adoption. This underscores the advantage of latent class modelling, which enables the identification of potential inherent resistance, a crucial factor in policy-making. By recognising these resistant segments, resources can be allocated more efficiently, avoiding investments in groups whose travel behaviours are unlikely to be influenced by external factors or policy interventions. Instead, efforts can be concentrated on segments that demonstrate greater openness to change, allowing for tailored policies that align with their characteristics and mode choice preferences.

For car users, although the Micromobility Resistant segment comprises 41.68\% and remains strongly committed to car travel, the other two segments exhibit a marked inclination toward alternative modes. In particular, the Multimodal Trip Supporters (36.70\%) demonstrate a moderate preference for an integrated travel approach, with approximately 47\% of their trips predicted to be made using public transport combined with shared micromobility. In contrast, the Personal Micromobility Lovers (21.62\%) show a more distinct and robust preference for personal micromobility, with over two-thirds of their trips predicted to be made using this mode.

Among public transport users, the segmentation reveals a clear divergence.The dominant group, identified as the Micromobility Adopters (77.92\%), shows a progressive attitude by distributing their mode choice almost equally between personal micromobility and an integrated travel option (42\% for each in prediction).  In contrast, the Micromobility Resistant group (22.08\%) remains heavily dependent on traditional public transport (89\%), with very limited uptake of EMM modes. This distinction suggests substantial opportunities to promote EMM adoption within the public transport segment, particularly among those already open to alternative modes. Interestingly, for those inclined towards EMM, the preference between personal and shared micromobility options appears evenly distributed, indicating a general openness to EMM regardless of its form.

\subsection{Class profile}
\begin{figure}[H] 
    \centering
    \begin{subfigure}[b]{0.9\textwidth}
        \centering
        \includegraphics[width=\textwidth, trim=0cm 5cm 0cm 0cm, clip]{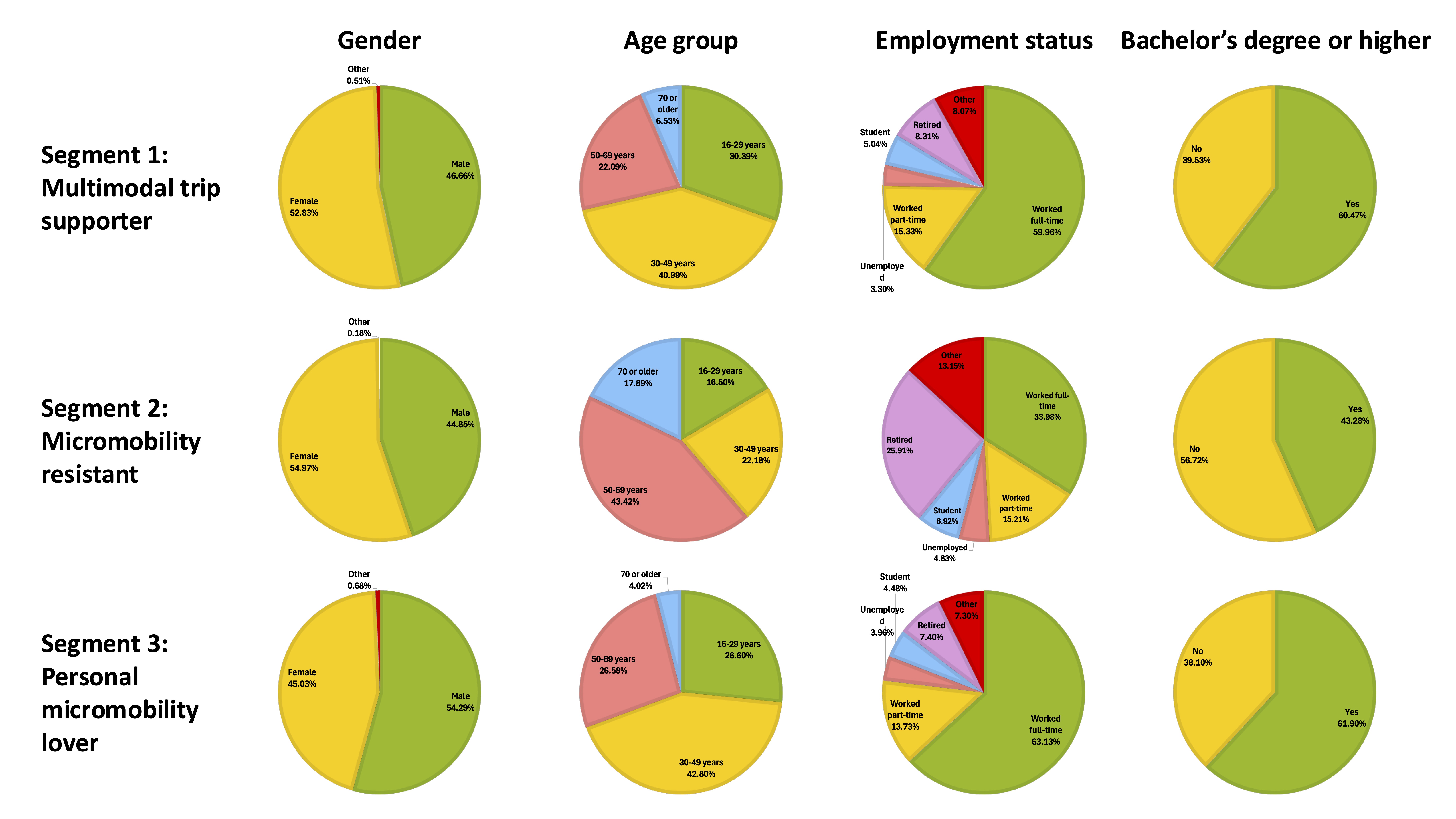}
    \end{subfigure}


    \begin{subfigure}[b]{0.9\textwidth}
        \centering
        \includegraphics[width=\textwidth, trim=0cm 5cm 0cm 0cm, clip]{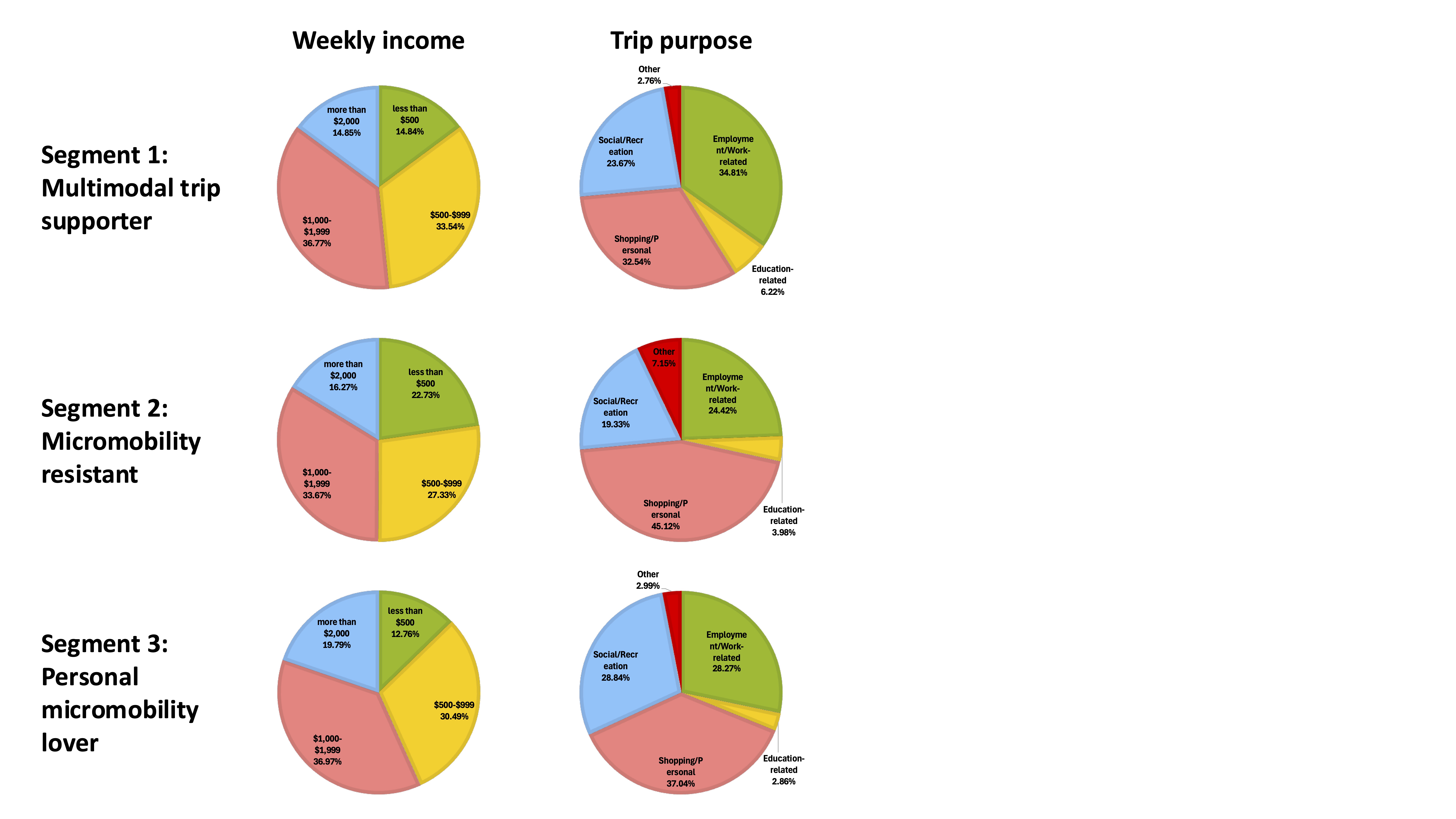}
    \end{subfigure}


    \begin{subfigure}[b]{0.9\textwidth}
        \centering
        \includegraphics[width=\textwidth, trim=0cm 5cm 0cm 0cm, clip]{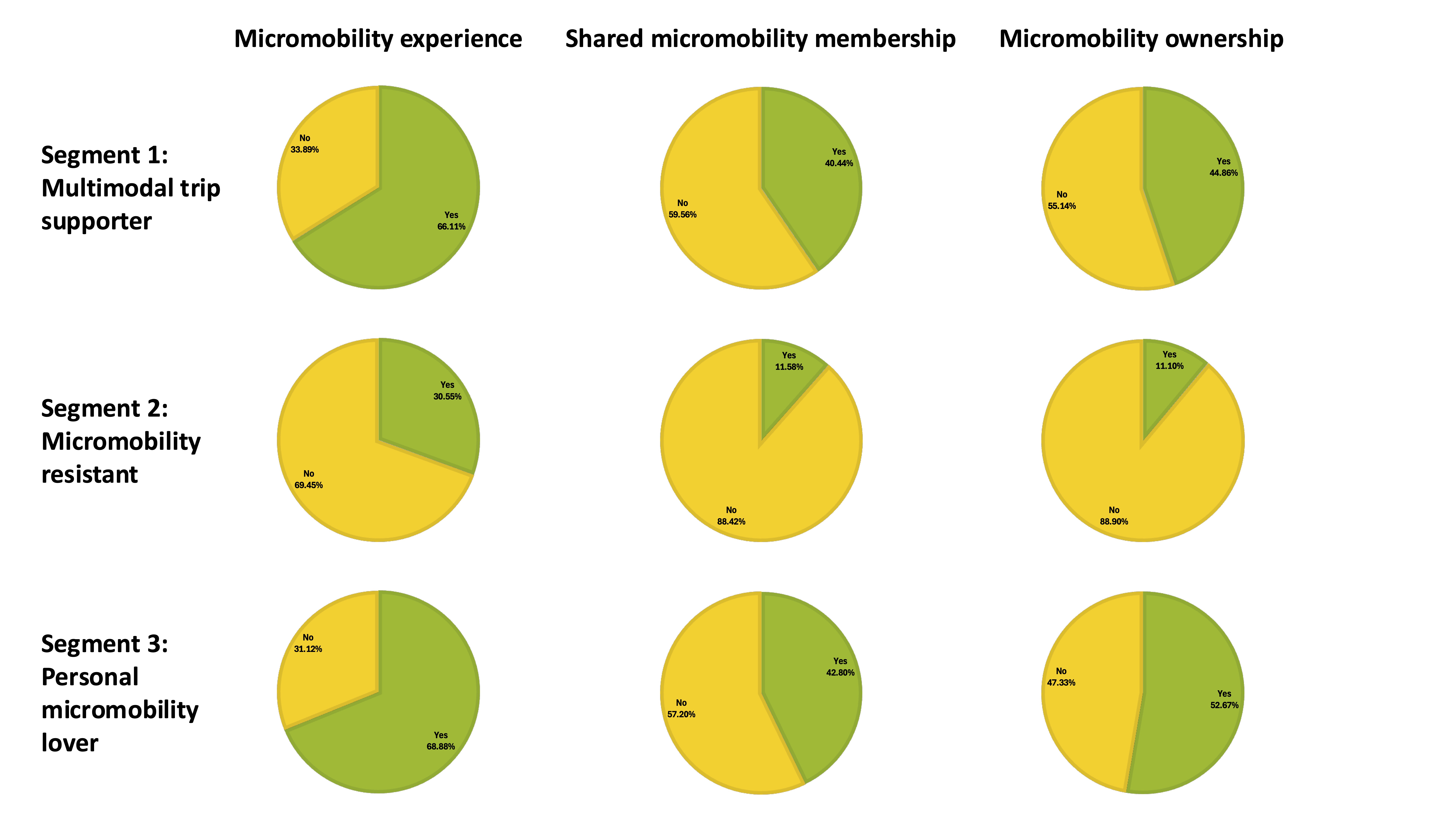}
    \end{subfigure}

    \caption{Characteristics of three segments of car users}
    \label{fig:car_user_profile}
\end{figure}

\begin{figure}[H] 
    \centering
    \begin{subfigure}[b]{0.9\textwidth}
        \centering
        \includegraphics[width=\textwidth, trim=0cm 50cm 0cm 5cm, clip]{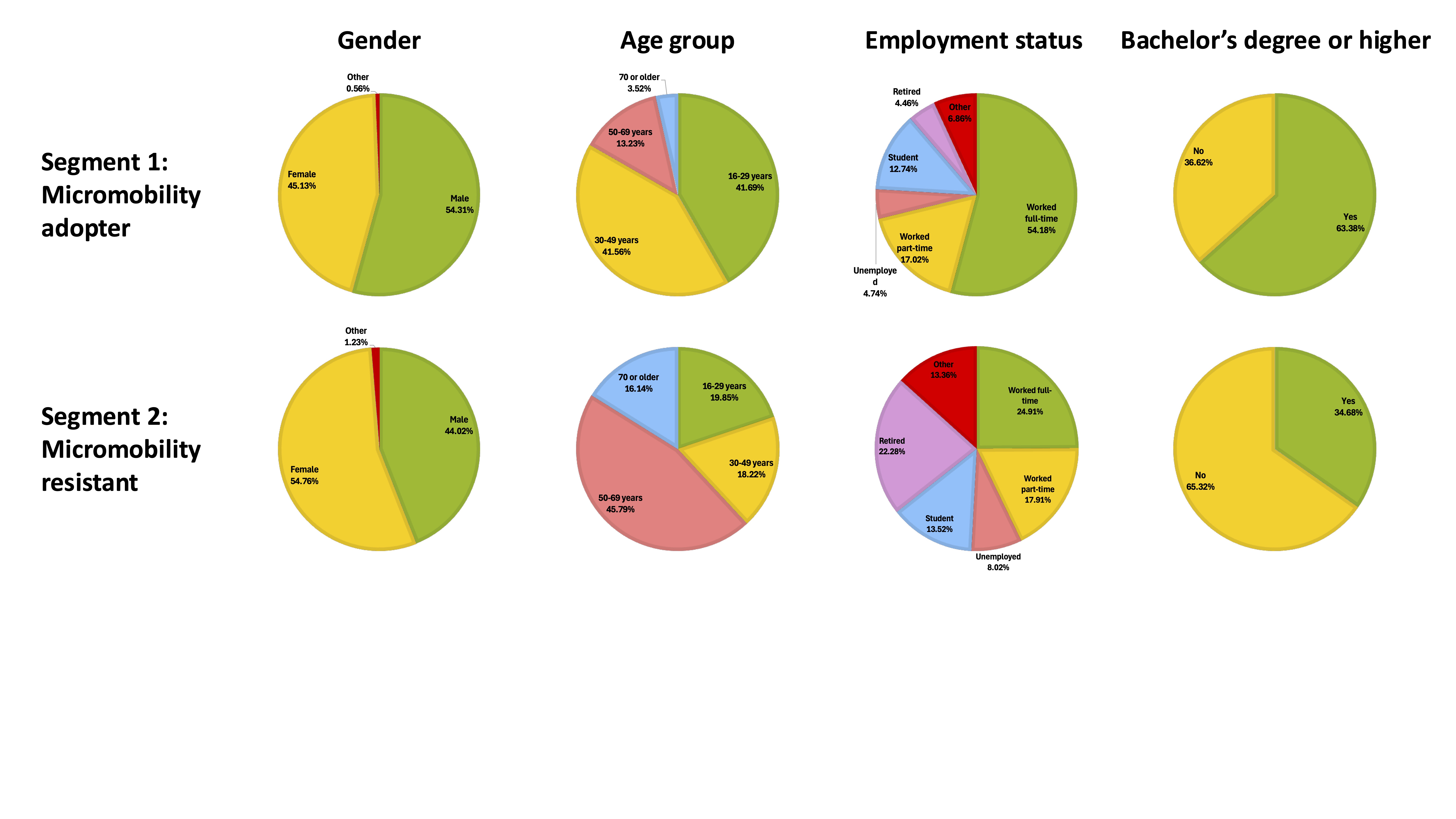}
    \end{subfigure}


    \begin{subfigure}[b]{0.9\textwidth}
        \centering
        \includegraphics[width=\textwidth, trim=0cm 50cm 0cm 5cm, clip]{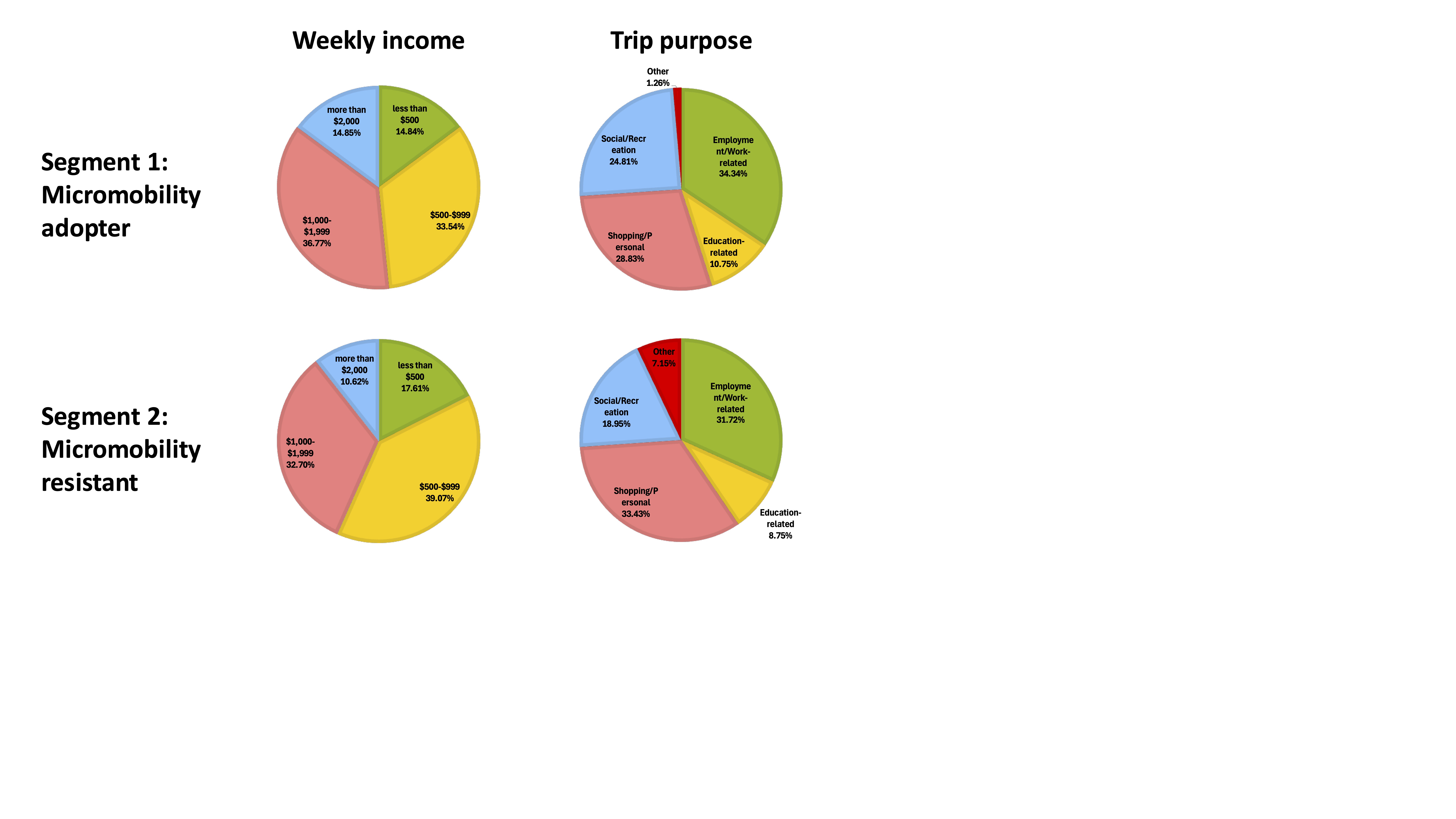}
    \end{subfigure}


    \begin{subfigure}[b]{0.9\textwidth}
        \centering
        \includegraphics[width=\textwidth, trim=0cm 50cm 0cm 5cm, clip]{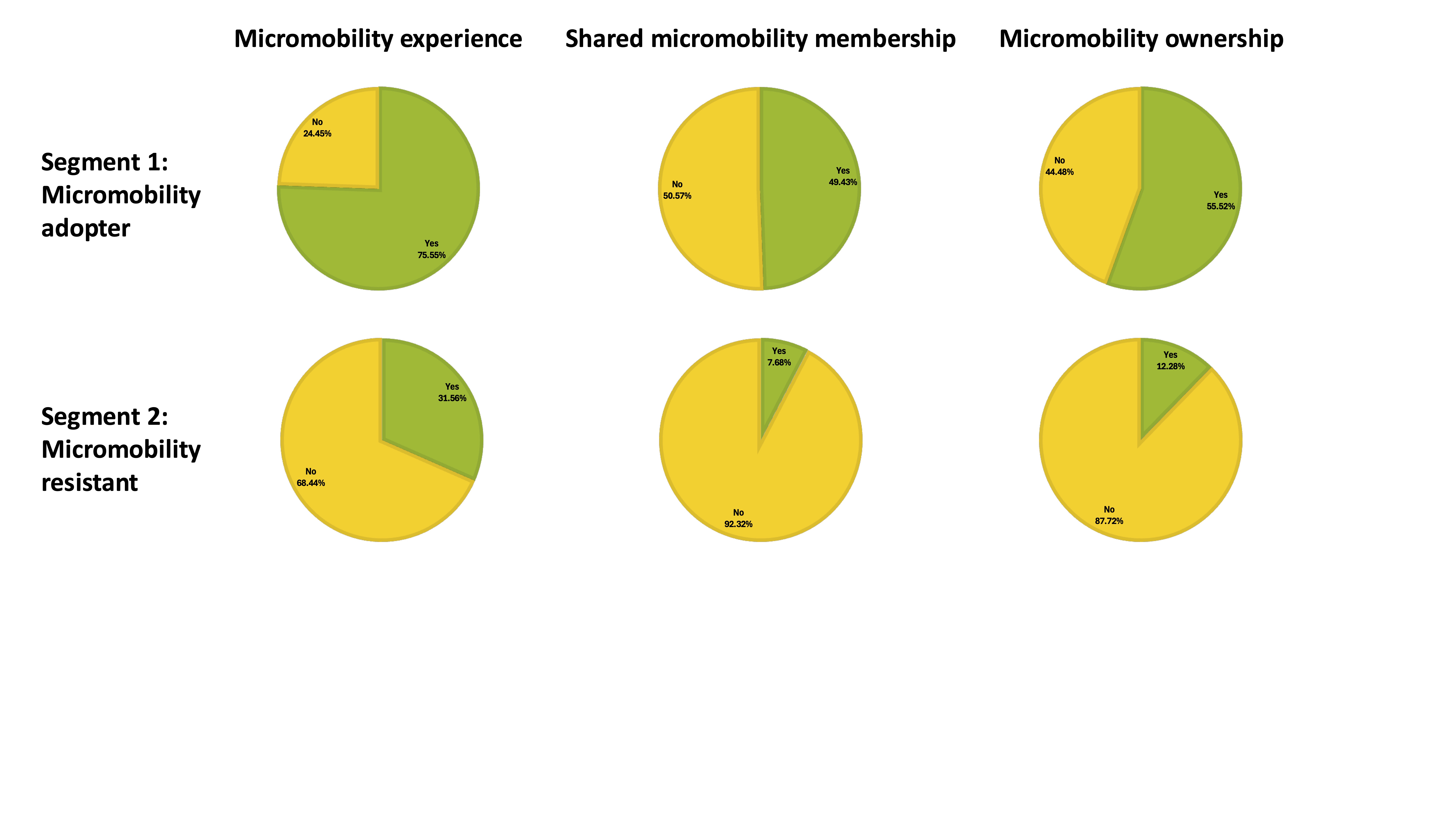}
    \end{subfigure}

    \caption{Characteristics of two segments of public transport users}
    \label{fig:pt_user_profile}
\end{figure}

Although the exact assignment of respondents to each class remains unknown, the class membership model enables estimation of individual-level posterior segment membership probabilities. These probabilities allow derivation of the proportional distribution of socio-demographic and trip-related characteristics for each segment. To better illustrate the sample profile for each class among both car users and public transport users, pie charts are presented in Fig.~\ref{fig:car_user_profile} and Fig.~\ref{fig:pt_user_profile}, respectively. For each segment of both car and public transport users, nine pie charts depict their distributions across gender, age, employment status, education level, income, and reference trip purpose, as well as three micromobility‐related attributes: whether they have prior experience using a micromobility device, hold a shared micromobility membership, or own a personal micromobility device.

Clear differences in characteristics across segments reveal significant heterogeneity within the sample. This heterogeneity enables the identification of specific groups whose mode choice behaviours are distinct, as captured by the latent class choice models. Consequently, such segmentation facilitates more targeted analysis and supports the formulation of tailored policy recommendations. In analysing willingness to change their original mode choice, common characteristics emerged among classes resistant to switching to EMM for their reference trip. In both groups of car and public transport users, individuals who showed resistance were often female, older, retired, had lower incomes and educational levels, used transportation primarily for shopping, and had notably less experience with micromobility.

Regarding car user segments, both the Multimodal Trip Supporters (Segment 1) and Personal Micromobility Lovers (Segment 3) exhibit markedly different profiles compared to the Micromobility Resistant group (Segment 2). Approximately 70\% of respondents in Segments 1 and 3 are under 49 years old, whereas only 38\% of those in Segment 2 fall within this age range. Furthermore, full-time employment is prevalent among the potential EMM adopters in Segments 1 and 3 (around 60\%), compared to only 33\% in the resistant group. A similar divergence is observed in educational attainment, with 60\% of potential EMM adopters holding a bachelor’s degree or higher, a stark contrast to the lower educational levels in Segment 2. In terms of trip purpose, shopping trips are more prominent among the Resistant segment, underscoring the continued importance of car cargo capacity. Notably, the Resistant group also shows a significant lack of prior EMM experience. Within the potential EMM user groups, further differences emerge: Personal Micromobility Lovers have a higher proportion of males (approximately 54\%) and a larger share of full-time workers, as well as a greater proportion of high-income individuals (with weekly incomes over \$2,000), compared to Multimodal Trip Supporters. Additionally, Multimodal Trip Supporters more frequently report employment- and education-related trips, while Personal Micromobility Lovers are more inclined toward social and shopping trips. An important trend is that over half (52.7\%) of Personal Micromobility Lovers already own an EMM device, highlighting the role of ownership in their mode choice behaviour.

Beyond the Resistant group (Segment 2), a Micromobility Adopter segment (Segment 1) emerges for public transport users, displaying notable heterogeneity in both demographics and mode choice behaviour. Similar to the car-user analysis, the distinction between EMM-leaning and Resistant segments among public transport users is even more pronounced here. In Segment 1, approximately 54\% of respondents are male—mirroring the female proportion in the Resistant group—and individuals under 49 years of age account for 83\%, compared to only 38\% in Segment 2. More than half (54.2\%) of the Micromobility Adopters work full-time, with fewer than 5\% retired, leading to higher income levels overall. By contrast, only 24.9\% of the Resistant segment are employed full-time, while 22.3\% are retired. Educational attainment also diverges sharply: 63\% of Segment 1 hold a bachelor’s degree or higher, whereas a similar share of Segment 2 do not. A notable distinction is also evident in attributes related to EMM, highlighting the considerable influence of EMM usage experience on their selection of modes.

\subsection{Elasticities}
While the sign of the estimated parameters indicates whether an attribute increases or decreases the likelihood of choosing a particular alternative, the coefficients themselves are not directly interpretable. Instead, elasticities capture the relative importance of each attribute by measuring the percentage change in choice probability for a 1\% change in the attribute level. These elasticities are computed by contrasting the model’s baseline predictions with its predictions after uniformly increasing the attribute value by 1\%. Table~\ref{tab:direct_elasticities} reports the direct elasticities (the effect of an attribute on the probability of choosing its own alternative), while Table~\ref{tab:cross_elasticities} presents the cross elasticities (the effect of a change in one alternative’s attribute on the probability of choosing competing alternatives). It is important to mention that only the attributes that showed significant results were further evaluated for their impact on elasticities and subsequently documented.

\begin{table}[htbp]
\centering
\caption{Summary of direct elasticities}
\label{tab:direct_elasticities}
\begin{threeparttable}
\subfloat[\textbf{Car Users (Three-Class LCCM)}]{
\begin{tabular}{llcccccc}
\toprule
\textbf{Class} & \textbf{Alternative} 
   & \textbf{Veh. TT} & \textbf{MM TT} & \textbf{Wait. T} 
   & \textbf{Walk. T} & \textbf{Run. Cost} & \textbf{Park. Cost} \\
\midrule
\multicolumn{8}{l}{\emph{Class 1 - Multimodal trip supporter}}\\
& PMM   & --          & -0.4705      & --       & --       & --       & -- \\
& PTSMM & -0.5593     & --           & -0.0830  & --       & --       & -- \\
& Car   & -0.5432     & --           & --       & -0.0982  & -0.2920  & -- \\
\\[-6pt]
\multicolumn{8}{l}{\emph{Class 2 - Micromobility resistant}}\\
& PMM   & --          & -1.6198      & --       & --       & --       & -- \\
& PTSMM & --          & --           & --       & --       & --       & -- \\
& Car   & -0.1742     & --           & --       & --       & -0.1409  & -0.0366 \\
\\[-6pt]
\multicolumn{8}{l}{\emph{Class 3 - Personal micromobility lover}}\\
& PMM   & --          & --           & --       & --       & --       & -- \\
& PTSMM & --          & --           & --       & --       & -0.4329  & -- \\
& Car   & --          & --           & --       & --       & --       & -- \\
\bottomrule
\end{tabular}
}\\
\subfloat[\textbf{Public Transport Users (Two-Class LCCM)}]{
\begin{tabular}{llcccccc}
\toprule
\textbf{Class} & \textbf{Alternative} 
   & \textbf{Veh. TT} & \textbf{MM TT} & \textbf{Wait. T} 
   & \textbf{Walk. T} & \textbf{Run. Cost} & \textbf{Park. Cost} \\
\midrule
\multicolumn{8}{l}{\emph{Class 1 - Micromobility adopter}}\\
& PMM   & --      & -0.4331 & --       & --       & --         & -- \\
& PTSMM & --      & --      & --       & --       & -0.1062    & -- \\
& PT    & --      & --      & --       & --       & -0.1816    & -- \\
\\[-6pt]
\multicolumn{8}{l}{\emph{Class 2 - Micromobility resistant}}\\
& PMM   & --      & --      & --       & --       & --         & -- \\
& PTSMM & --      & --      & --       & --       & --         & -- \\
& PT    & --      & --      & --       & --       & --         & -- \\
\bottomrule
\end{tabular}
}
\end{threeparttable}
\end{table}

\begin{table}[htbp]
\centering
\caption{Cross elasticities of the latent segments}
\label{tab:cross_elasticities}
\resizebox{0.85\textwidth}{!}{%
\begin{threeparttable}
\subfloat[\textbf{Car Users (Three-Class LCCM)}]{
\begin{tabular}{l rrr rrr rrr}
\toprule
& \multicolumn{3}{c}{\makecell{\textbf{Segment 1}\\ Multimodal trip supporter} } 
& \multicolumn{3}{c}{\textbf{\makecell{\textbf{Segment 2} \\ Micromobility resistant}}} 
& \multicolumn{3}{c}{\makecell{\textbf{Segment 3}\\ Personal micromobility lover}} \\
\cmidrule(lr){2-4}\cmidrule(lr){5-7}\cmidrule(lr){8-10}
\textbf{Attribute} 
& \textbf{PMM} & \textbf{PTSMM} & \textbf{Car}
& \textbf{PMM} & \textbf{PTSMM} & \textbf{Car}
& \textbf{PMM} & \textbf{PTSMM} & \textbf{Car} \\
\midrule
PT travel time        &  0.4881 &   --    &  0.5049 
                     &    --    &   --    &    --   
                     &    --    &   --    &   --    \\
Car travel time       &  0.2136 & 0.2273  &   --    
                     & 1.5692   & 1.7091  &   --    
                     &    --    &   --    &   --    \\
PMM travel time       &    --   & 0.1482  & 0.1403  
                     &    --    & 0.2359  & 0.0848  
                     &    --    &   --    &   --    \\
PT waiting time       &  0.0748 &   --    & 0.0736  
                     &    --    &   --    &   --    
                     &    --    &   --    &   --    \\
Car walking time      &  0.0400 & 0.0406  &   --    
                     &    --    &   --    &   --    
                     &    --    &   --    &   --    \\
PMM running cost      &    --   &   --    &   --    
                     &    --    &   --    &   --    
                     &    --    &   --    &   --    \\
PTSMM running cost    &    --   &   --    &   --    
                     &    --    &   --    &   --    
                     & 0.0921   &   --    & 0.0773  \\
Car running cost      &  0.1127 & 0.1236  &   --    
                     & 1.2679   & 1.3892  &   --    
                     &    --    &   --    &   --    \\
Car parking cost      &    --   &   --    &   --    
                     & 0.3229   & 0.3724  &   --    
                     &    --    &   --    &   --    \\
\bottomrule
\end{tabular}
}

\vspace{1em}

\subfloat[\textbf{Public Transport Users (Two-Class LCCM)}]{
\begin{tabular}{l rrr rrr}
\toprule
& \multicolumn{3}{c}{\textbf{Segment 1}} 
& \multicolumn{3}{c}{\textbf{Segment 2}} \\
\cmidrule(lr){2-4}\cmidrule(lr){5-7}
\textbf{Attribute} 
& \textbf{PMM} & \textbf{PTSMM} & \textbf{PT}
& \textbf{PMM} & \textbf{PTSMM} & \textbf{PT} \\
\midrule
PT travel time        &   --   & -- & --
                     &   --   &   --   &  --    \\
PMM travel time       &   --   &   0.3222   &   0.2928   
                     &   --   &   --   &   --   \\
SMM travel time       &   --   &   --   &   --   
                     &   --   &   --   &   --   \\
PTSMM waiting time    &   --   &   --   &   --   
                     &   --   &   --   &   --   \\
PT waiting time       &   --   &   --   &   --   
                     &   --   &   --   &   --   \\
PT walking time       &   --   &   --   &   --   
                     &   --   &   --   &   --   \\
PMM running cost      &   --   &   --   &   --   
                     &   --   &   --   &   --   \\
PTSMM running cost    & 0.0763 &   --   & 0.0736 
                     &   --   &   --   &   --   \\
PT running cost       & 0.0344 & 0.0363 &   --   
                     &   --   &   --   &   --   \\
\bottomrule
\end{tabular}
}
\end{threeparttable}
}
\end{table}

For car users, elasticity estimates for the Resistant group provide clear insights into the relative effects of various attributes, thereby helping to prioritise interventions that can efficiently facilitate a shift toward EMM. Notably, Multimodal Trip Supporters (Segment 1) are sensitive to a range of factors, whereas the choices of Personal Micromobility Lovers (Segment 3) are barely affected by these changes. Direct elasticity estimates indicate that a 1\% increase in micromobility travel time results in more than a 1.6\% decrease in the probability of choosing personal micromobility for the Resistant group, in contrast to a much lower sensitivity among Multimodal Trip Supporters, who exhibit a direct elasticity of -0.47. Moreover, the Resistant group exhibits cross elasticities greater than one for EMM alternatives, underscoring the potential for mode shift driven by these attributes. Specifically, a 1\% increase in car travel time corresponds to a 1.57\% and 1.71\% increase in the probability of selecting personal micromobility and the integrated multimodal alternative, respectively. Likewise, a 1\% increase in car travel cost leads to increases of 1.27\% and 1.39\% in the probability of choosing these alternatives. Secondary factors—such as waiting time, walking time, and parking cost—have relatively minor impacts on mode choice probability. In general, travel cost factors show a comparatively lesser effect in elasticity estimates when compared to travel time factors.

The elasticities for public transport users differ markedly from those observed for car users, largely due to the limited number of significant attributes in the LCCM for this group. Consequently, there are fewer meaningful elasticity estimates, and their magnitudes are generally smaller. Notably, no meaningful elasticities were identified for the Resistant segment among public transport users. For the Micromobility Adopter segment (Segment 1), a 1\% increase in personal micromobility travel time is associated with a 0.4\% decrease in the probability of choosing that alternative, while it leads to increases of 0.32\% and 0.29\% in the probabilities of selecting the public transport with shared micromobility option and conventional public transport, respectively. Furthermore, if Segment 1 tends to favour personal micromobility over the other options, then increasing the travel costs associated with the multimodal and public transport alternatives can further shift choice toward personal micromobility—with a 1\% increase in cost resulting in approximately 0.08\% and 0.03\% increases in the probability of choosing personal micromobility, respectively.

\subsection{Geographical distribution of sample across classes}
\begin{figure}[h!] 
    \centering
    \begin{subfigure}[b]{0.32\textwidth}
        \centering
        \includegraphics[width=\textwidth, trim=0cm 0cm 0cm 1cm, clip]{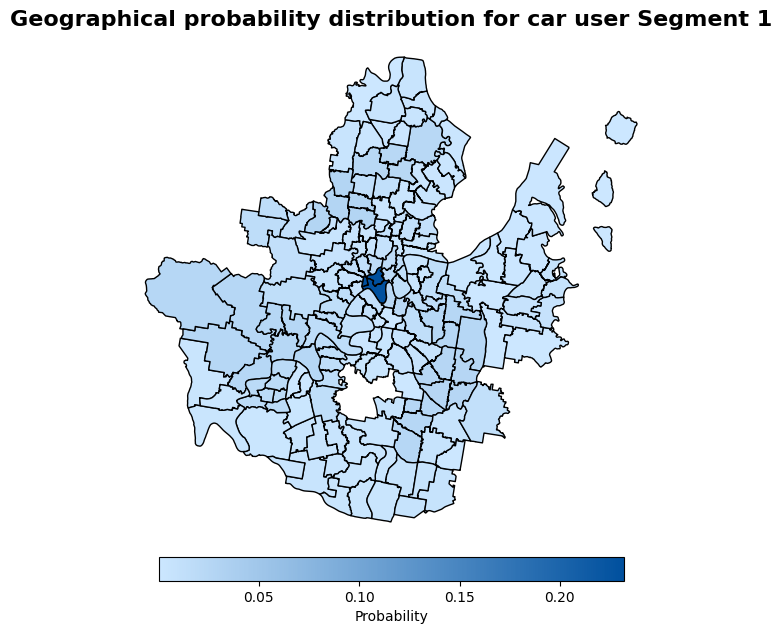}
        \caption{Segment 1}
        \label{fig:car_geo_1}
    \end{subfigure}
    \hfill
    \begin{subfigure}[b]{0.32\textwidth}
        \centering
        \includegraphics[width=\textwidth, trim=0cm 0cm 0cm 1cm, clip]{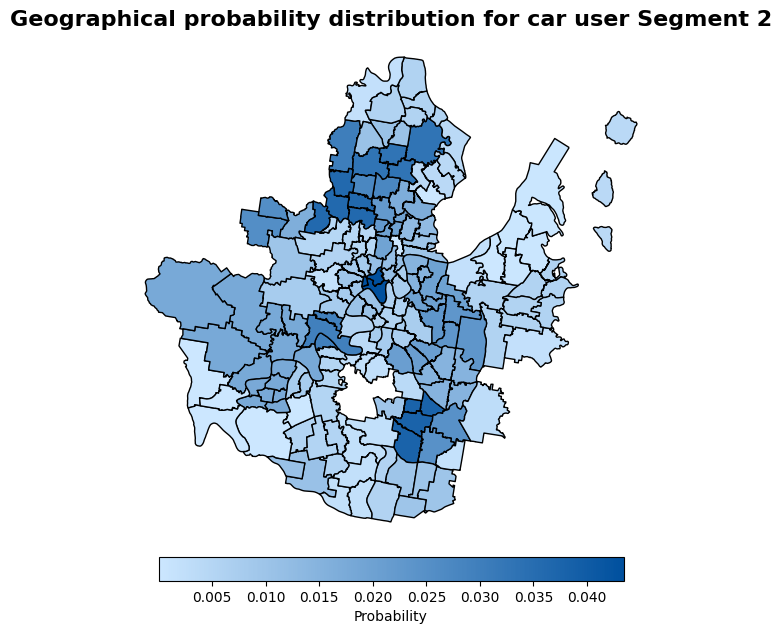}
        \caption{Segment 2}
        \label{fig:car_geo_2}
    \end{subfigure}
    \hfill
    \begin{subfigure}[b]{0.32\textwidth}
        \centering
        \includegraphics[width=\textwidth, trim=0cm 0cm 0cm 1cm, clip]{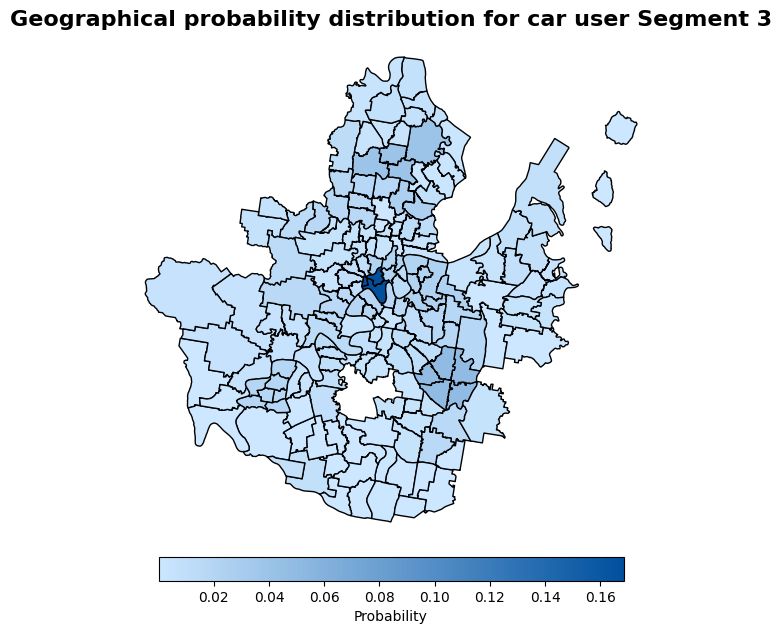}
        \caption{Segment 3}
        \label{fig:car_geo_3}
    \end{subfigure}

    \caption{Living suburbs probability distribution of car users}
    \label{fig:car_user_spatial}
\end{figure}

\begin{figure}[h!] 
    \centering
    \begin{subfigure}[b]{0.48\textwidth}
        \centering
        \includegraphics[width=\textwidth, trim=0cm 0cm 0cm 1cm, clip]{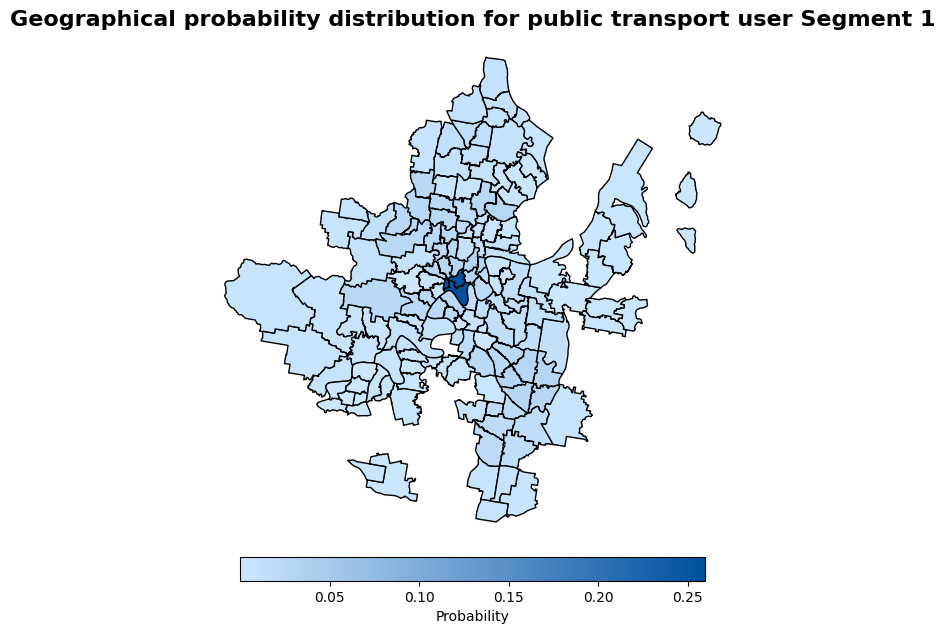}
        \caption{Segment 1}
        \label{fig:pt_geo_1}
    \end{subfigure}
    \hfill
    \begin{subfigure}[b]{0.48\textwidth}
        \centering
        \includegraphics[width=\textwidth, trim=0cm 0cm 0cm 1cm, clip]{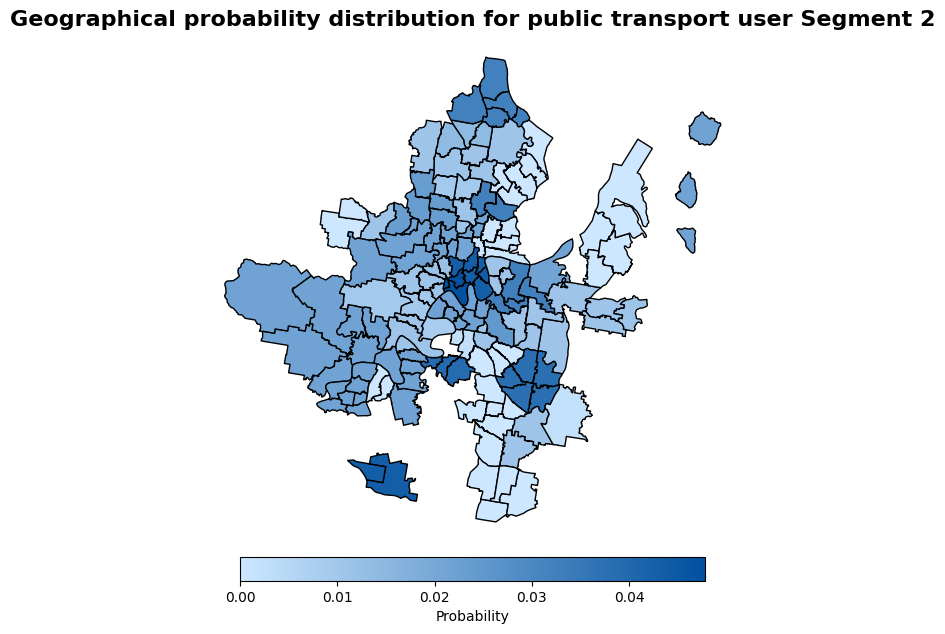}
        \caption{Segment 2}
        \label{fig:pt_geo_2}
    \end{subfigure}

    \caption{Living suburbs probability distribution of public transport users}
    \label{fig:pt_user_spatial}
\end{figure}

To better understand the factors contributing to differences in mode choice behaviour across latent classes, the residential area data were analysed to assess the potential impact of the built environment. Information on respondents' suburbs of residence was collected, and using a similar procedure as that for calculating the probability of segment characteristics, the probability distribution of residential areas for each latent class was derived. Figure~\ref{fig:car_user_spatial} and Figure~\ref{fig:pt_user_spatial} illustrate the density of residential suburbs for each segment among both car users and public transport users, where darker areas indicate a higher likelihood that respondents from that class reside there. It is important to note that each subfigure uses a different colour scale to depict segment‐specific probabilities. Consequently, although the city centres may appear similarly shaded across figures, the actual probability values differ considerably. In particular, the similarly dark shading in the central city areas for EMM‐leaning segments actually reflects higher probability densities than those of the resistant segments (Segment 2 for both car and public transport). 

The dark area in the center of the figures represents Brisbane’s city center. It is evident that segments inclined to EMM usage (Segment 1 and 3 for car users; Segment 1 for public transport users) are predominantly concentrated in the urban core, while the micromobility-resistant segments (Segment 2 for both car and public transport users) are dispersed across suburban areas. This spatial disparity is likely driven by factors such as higher population densities, shorter travel distances, and more extensive micromobility infrastructure in urban areas, which encourage EMM adoption. In contrast, suburban environments, with their longer commutes and greater reliance on carrying goods or people, tend to foster resistance to EMM. These differences in residential distribution reveal significant heterogeneity in living areas and help pinpoint where potential mode shift interventions could be most effectively targeted.

\section{Discussions of results}

Results from the LCCMs estimated in this study reveal a strong potential for car and public transport users to shift their mode choice toward EMM alternatives. At the same time, considerable heterogeneity exists across different latent classes, indicating that tailored policy interventions are necessary to maximise EMM adoption. By analysing the choice behaviour preferences and demographic characteristics of EMM-leaning segments, targeted policy schemes can be designed to enhance cost efficiency and increase the rate of EMM adoption, thereby alleviating traffic congestion through a mode shift.

\subsection{Car users}
Car users in this study are classified into three distinct segments: one resistant group (41.68\%) and two segments that show potential for shifting away from car use, together accounting for nearly 60\% of the sample.

\paragraph{Multimodal Trip Supporters (36.7\%)} demonstrate a clear preference for using public transport integrated with shared micromobility over both personal micromobility and continued car travel. Their willingness to adopt EMM alternatives is significantly influenced by weather conditions, particularly rainy weather, which serves as a major deterrent. While travel time and cost hold secondary effects, their mode choice is sensitive to a wider range of factors, including in-vehicle, waiting, walking time, and car running cost. Their choice behaviour is relatively flexible and easily influenced by trip-specific factors, making this segment a key target for policy interventions. While this flexibility presents significant potential for shifting toward EMM-based travel, it also poses a challenge: their usage of EMM alternatives is not firmly established and could diminish over time if external conditions—such as convenience and reliability—are not continuously optimised.

\paragraph{Personal Micromobility Lovers (21.62\%)} is far less affected by travel time or cost considerations, with rainy weather being the primary deterrent to EMM use, highlights the importance of cycling infrastructure, particularly rain-protected bike lanes. This group's clear preference for personal micromobility is further highlighted by the fact that the running cost of a multimodal option is the only significant influence on their choices. Despite the effect's minor magnitude, it indicates a tendency toward personal micromobility rather than a multimodal journey. Nevertheless, it also indicates the likelihood that the multimodal option might be embraced by decreasing its operational expenses, potentially through a bundled discount strategy. Given that this group is already inclined toward EMM, policies should concentrate on maintaining and enhancing their user experience to ensure long-term adoption.

\subsection{Public transport users}
Public transport users show an even stronger tendency to shift toward EMM-related modes, with around 78\% of this group predicted to choose either personal micromobility or a multimodal option that combines public transport with shared micromobility. Similar to car users, public transport users are significantly affected by rainy weather, which presents a major barrier to EMM adoption. Ensuring weather protection in infrastructure design is crucial to support their transition. Unlike car users, public transport users are more concerned with running costs than travel time. This suggests that pricing incentives, such as discounted multimodal fare bundles, could be an effective strategy to encourage mode shifts.


\subsection{Policy implications}
\paragraph{Investment in dedicated micromobility infrastructure} should be prioritised to encourage EMM adoption to provide weather protection and travel time reduction. Since both EMM-leaning car and public transport users are highly sensitive to rainy weather, installing covered bike lanes, sheltered docking stations, and rain-resistant pathways can significantly enhance EMM usability. Additionally, reducing travel time through dedicated bike lanes, optimized traffic signals for non-motorized users, and seamless integration with transit hubs—along with reducing public transport service waiting times, particularly for the Multimodal Trip Supporter group among car users—can enhance the appeal and efficiency of EMM. These improvements can significantly improve the riding experience for EMM-leaning groups, making mode shifts more sustainable and stable over time. Moreover, initial investments should be prioritised in urban areas, where EMM-leaning users are densely concentrated in city centres, maximising the impact of infrastructure improvements.

\paragraph{Shared micromobility pricing discounts} can be a key strategy to encourage multimodal EMM adoption, particularly among public transport users who prioritise cost over travel time, as well as the Personal Micromobility Lover segment among car users, helping this group explore multimodal alternatives. Bundled public transport and shared micromobility discounts, such as offering free or reduced-cost shared micromobility rides when combined with public transit, can effectively incentivise multimodal trips. Additionally, promoting shared micromobility memberships through loyalty programs or government subsidies can further lower financial barriers, making EMM a more accessible and attractive option for daily commuters. To maximise the effectiveness of these discounts, policies should target urban public transport users under 49 years old, full-time workers, and those with a bachelor’s degree or higher, as these groups exhibit the highest potential for adopting EMM.

\paragraph{EMM experience as a key factor in adoption} is evident from the class profile analysis. Policies promoting hands-on experience with EMM can be a powerful tool for shifting resistant users toward adoption, aligning with findings from prior studies (e.g., \citep{chen2023exploring}). Targeted experience-based campaigns should focus on specific demographic groups to maximise their impact. For younger, higher-income male car users, offering free trials of personal micromobility, such as e-bike or e-scooter test rides, can significantly increase their willingness to transition to personal micromobility use. Similarly, for public transport users interested in shared micromobility, providing free or discounted access to shared micromobility services—particularly in city centre areas where these users are concentrated—can offer valuable hands-on exposure, encouraging long-term use and reinforcing multimodal travel habits.

\section{ Conclusion}
EMM has been increasingly recognised as a sustainable, flexible, and cost-effective alternative to private car travel, offering a promising solution to address the ‘first and last mile’ challenge in public transport integration. However, existing research has often overlooked the heterogeneity in user preferences and the varying factors influencing EMM adoption. This study provides valuable insights into the mode choice behaviours of car and public transport users, analysing their preferences for EMM-related alternatives using a Latent Class Choice Model (LCCM) approach. Based on survey data from 1,672 participants in Brisbane, our findings identify distinct user segments with varying levels of willingness to adopt EMM.

The LCCM predictions indicate that nearly 60\% of car users and 78\% of public transport users demonstrate some degree of willingness to transition to EMM-related alternatives, particularly under favourable conditions such as reduced travel time, improved infrastructure, and financial incentives. However, key deterrents, such as adverse weather conditions, remain significant barriers to widespread adoption. By integrating elasticity analysis, class profiles, and spatial distributions, this study proposes practical policy recommendations tailored to different user segments. A customised policy framework, designed to maximise mode shift from traditional transport modes to EMM, can ensure a cost-efficient transition toward sustainable urban mobility.

\subsection{Limitations and future research directions}
Despite providing a comprehensive analysis of mode choice behaviours, this study has certain limitations that highlight directions for future research. As a stated-preference survey study, it is subject to hypothetical bias, despite efforts to minimise this effect. Future research could integrate real-world behavioural data with stated preference surveys to gain deeper insights into actual mode choice decisions. Additionally, exploring long-term mode choice behaviour in real-world settings could help account for dynamic factors, such as evolving transport policies, infrastructure changes, and seasonal variations, leading to a more accurate assessment of EMM promotion strategies.

Moreover, this study was conducted in Brisbane, a city characterised by a sunny climate and hilly terrain, which may limit the generalisability of the findings to other urban contexts. Future research should explore cross-regional studies to validate the model’s applicability in different metropolitan areas with diverse environmental and infrastructural conditions. Furthermore, some factors, such as congestion levels and cycling lane proportions, were found to be statistically insignificant, potentially due to participants’ limited perception in an online survey setting. Incorporating immersive tools, such as virtual reality simulations, could help generate more realistic choice scenarios, allowing for a more reliable observation of decision-making behaviours and improving the accuracy of mode choice modelling.

Although LCCM can effectively extract the heterogeneity over sample, in real-world scenarios, the assumption of rational behavior is often violated, as psychological factors, including attitudes, perceptions, and habits, significantly influence individual decision-making, alongside rational factors \citep{mcfadden1986choice}. To integrate psychological dimensions, a hybrid choice model (HCM) framework—also known as the integrated choice and latent variable (ICLV) model—has been widely applied. This framework incorporates both a latent variable model and a discrete choice model (including LCCM) to capture preference heterogeneity across distinct groups \citep{ben2002integration}. By addressing these research gaps, future studies can further enhance the understanding of EMM adoption dynamics and support the development of more effective policies to encourage sustainable mobility transitions.



\begin{small}
\begin{sloppypar} 
\bibliographystyle{authordate1} 

\setlength{\bibsep}{0pt}

\bibliography{References}

\begin{thebibliography}{}

\bibitem[\protect\citename{Asgari {\em et~al.}, }2018]{asgari2018stated}
Asgari, Hamidreza, Jin, Xia, \& Corkery, Terrence. 2018.
\newblock A stated preference survey approach to understanding mobility choices in light of shared mobility services and automated vehicle technologies in the US.
\newblock {\em Transportation Research Record}, {\bf 2672}(47), 12--22.

\bibitem[\protect\citename{Baek {\em et~al.}, }2021]{baek2021electric}
Baek, Kwangho, Lee, Hyukseong, Chung, Jin-Hyuk, \& Kim, Jinhee. 2021.
\newblock Electric scooter sharing: How do people value it as a last-mile transportation mode?
\newblock {\em Transportation Research Part D: Transport and Environment}, {\bf 90}, 102642.

\bibitem[\protect\citename{Ben-Akiva {\em et~al.}, }2002]{ben2002integration}
Ben-Akiva, Moshe, Walker, Joan, Bernardino, Adriana~T, Gopinath, Dinesh~A, Morikawa, Taka, \& Polydoropoulou, Amalia. 2002.
\newblock Integration of choice and latent variable models.
\newblock {\em Perpetual motion: Travel behaviour research opportunities and application challenges}, {\bf 2002}, 431--470.

\bibitem[\protect\citename{Cao {\em et~al.}, }2021]{cao2021scooter}
Cao, Zhejing, Zhang, Xiaohu, Chua, Kelman, Yu, Honghai, \& Zhao, Jinhua. 2021.
\newblock E-scooter sharing to serve short-distance transit trips: A Singapore case.
\newblock {\em Transportation research part A: policy and practice}, {\bf 147}, 177--196.

\bibitem[\protect\citename{Census, }2021]{ABS2021}
Census, 2021~Australian. 2021.
\newblock {\em ABS2021}.
\newblock \url{https://www.abs.gov.au/census/find-census-data/quickstats/2021/3}.

\bibitem[\protect\citename{Chen {\em et~al.}, }2023]{chen2023exploring}
Chen, Ching-Fu, Fu, Chiang, \& Siao, Pei-Ya. 2023.
\newblock Exploring electric moped sharing preferences with integrated choice and latent variable approach.
\newblock {\em Transportation Research Part D: Transport and Environment}, {\bf 121}, 103837.

\bibitem[\protect\citename{ChoiceMetrics, }2012]{choicemetrics2012ngene}
ChoiceMetrics, C. 2012.
\newblock Ngene 1.1. 1 user manual \& reference guide.
\newblock {\em Sydney, Australia}.

\bibitem[\protect\citename{Clewlow, }2018]{clewlow2018micro}
Clewlow, R. 2018.
\newblock The Micro-Mobility Revolution [WWW Document].
\newblock {\em URL https://medium. com/populus-ai/the-micro-mobility-revolution-95e396db3754 (accessed 7.10. 20)}.

\bibitem[\protect\citename{Cui \& Zhang, }2024]{cui2024integration}
Cui, Can, \& Zhang, Yu. 2024.
\newblock Integration of Shared Micromobility into Public Transit: A Systematic Literature Review with Grey Literature.
\newblock {\em Sustainability}, {\bf 16}(9), 3557.

\bibitem[\protect\citename{Curtale {\em et~al.}, }2021]{curtale2021understanding}
Curtale, Riccardo, Liao, Feixiong, \& van~der Waerden, Peter. 2021.
\newblock Understanding travel preferences for user-based relocation strategies of one-way electric car-sharing services.
\newblock {\em Transportation Research Part C: Emerging Technologies}, {\bf 127}, 103135.

\bibitem[\protect\citename{de~Dios~Ort{\'u}zar \& Willumsen, }2024]{de2024modelling}
de~Dios~Ort{\'u}zar, Juan, \& Willumsen, Luis~G. 2024.
\newblock {\em Modelling transport}.
\newblock John wiley \& sons.

\bibitem[\protect\citename{Eom {\em et~al.}, }2023]{eom2023exploring}
Eom, Jin~Ki, Lee, Kwang-Sub, \& Lee, Jun. 2023.
\newblock Exploring micromobility mode preferences for last-mile trips from subway stations.
\newblock {\em Journal of Public Transportation}, {\bf 25}, 100054.

\bibitem[\protect\citename{Eszterg{\'a}r-Kiss {\em et~al.}, }2022]{esztergar2022assessment}
Eszterg{\'a}r-Kiss, Domokos, Tordai, D{\'a}niel, \& Lizarraga, Julio C~Lopez. 2022.
\newblock Assessment of travel behavior related to e-scooters using a stated preference experiment.
\newblock {\em Transportation Research Part A: Policy and Practice}, {\bf 166}, 389--405.

\bibitem[\protect\citename{Fishman {\em et~al.}, }2015]{fishman2015factors}
Fishman, Elliot, Washington, Simon, Haworth, Narelle, \& Watson, Angela. 2015.
\newblock Factors influencing bike share membership: An analysis of Melbourne and Brisbane.
\newblock {\em Transportation research part A: policy and practice}, {\bf 71}, 17--30.

\bibitem[\protect\citename{Fitt \& Curl, }2019]{fitt2019scooter}
Fitt, Helen, \& Curl, Angela. 2019.
\newblock E-scooter use in New Zealand: Insights around some frequently asked questions.
\newblock {\em University of Canterbury: Christchurch, New Zealand}.

\bibitem[\protect\citename{Greene \& Hensher, }2003]{greene2003latent}
Greene, William~H, \& Hensher, David~A. 2003.
\newblock A latent class model for discrete choice analysis: contrasts with mixed logit.
\newblock {\em Transportation Research Part B: Methodological}, {\bf 37}(8), 681--698.

\bibitem[\protect\citename{Guo \& Zhang, }2021]{guo2021understanding}
Guo, Yujie, \& Zhang, Yu. 2021.
\newblock Understanding factors influencing shared e-scooter usage and its impact on auto mode substitution.
\newblock {\em Transportation research part D: transport and environment}, {\bf 99}, 102991.

\bibitem[\protect\citename{Hess \& Palma, }2019]{hess2019apollo}
Hess, Stephane, \& Palma, David. 2019.
\newblock Apollo: A flexible, powerful and customisable freeware package for choice model estimation and application.
\newblock {\em Journal of choice modelling}, {\bf 32}, 100170.

\bibitem[\protect\citename{Jaber {\em et~al.}, }2023]{jaber2023preferences}
Jaber, Ahmed, Hamadneh, Jamil, \& Csonka, Bálint. 2023.
\newblock The Preferences of Shared Micro-Mobility Users in Urban Areas.
\newblock {\em IEEE Access}, {\bf 11}, 74458--74472.

\bibitem[\protect\citename{Johnson {\em et~al.}, }2023]{johnson2023impacts}
Johnson, Nicholas, Fitch-Polse, Dillon~T, \& Handy, Susan~L. 2023.
\newblock Impacts of e-bike ownership on travel behavior: Evidence from three northern California rebate programs.
\newblock {\em Transport policy}, {\bf 140}, 163--174.

\bibitem[\protect\citename{Lazarsfeld, }1968]{lazarsfeld1968latent}
Lazarsfeld, Paul~Felix. 1968.
\newblock Latent structure analysis.
\newblock {\em (No Title)}.

\bibitem[\protect\citename{Louviere {\em et~al.}, }2000]{louviere2000stated}
Louviere, Jordan~J, Hensher, David~A, \& Swait, Joffre~D. 2000.
\newblock {\em Stated choice methods: analysis and applications}.
\newblock Cambridge university press.

\bibitem[\protect\citename{McFadden, }1974]{mcfadden1974measurement}
McFadden, Daniel. 1974.
\newblock The measurement of urban travel demand.
\newblock {\em Journal of public economics}, {\bf 3}(4), 303--328.

\bibitem[\protect\citename{McFadden, }1986]{mcfadden1986choice}
McFadden, Daniel. 1986.
\newblock The choice theory approach to market research.
\newblock {\em Marketing science}, {\bf 5}(4), 275--297.

\bibitem[\protect\citename{Merlin {\em et~al.}, }2021]{merlin2021segment}
Merlin, Louis~A, Yan, Xiang, Xu, Yiming, \& Zhao, Xilei. 2021.
\newblock A segment-level model of shared, electric scooter origins and destinations.
\newblock {\em Transportation Research Part D: Transport and Environment}, {\bf 92}, 102709.

\bibitem[\protect\citename{Milakis {\em et~al.}, }2020]{milakis2020micro}
Milakis, Dimitris, Gedhardt, Laura, Ehebrecht, Daniel, \& Lenz, Barbara. 2020.
\newblock Is micro-mobility sustainable? An overview of implications for accessibility, air pollution, safety, physical activity and subjective wellbeing.
\newblock {\em Handbook of sustainable transport},  180--189.

\bibitem[\protect\citename{Montes {\em et~al.}, }2023]{montes2023shared}
Montes, Alejandro, Ger{\v{z}}inic, Nejc, Veeneman, Wijnand, van Oort, Niels, \& Hoogendoorn, Serge. 2023.
\newblock Shared micromobility and public transport integration-A mode choice study using stated preference data.
\newblock {\em Research in Transportation Economics}, {\bf 99}, 101302.

\bibitem[\protect\citename{Nikiforiadis {\em et~al.}, }2021]{nikiforiadis2021analysis}
Nikiforiadis, Andreas, Paschalidis, Evangelos, Stamatiadis, Nikiforos, Raptopoulou, Alexandra, Kostareli, Athanasia, \& Basbas, Socrates. 2021.
\newblock Analysis of attitudes and engagement of shared e-scooter users.
\newblock {\em Transportation research part D: transport and environment}, {\bf 94}, 102790.

\bibitem[\protect\citename{Reck {\em et~al.}, }2021]{reck2021explaining}
Reck, Daniel~J, Haitao, He, Guidon, Sergio, \& Axhausen, Kay~W. 2021.
\newblock Explaining shared micromobility usage, competition and mode choice by modelling empirical data from Zurich, Switzerland.
\newblock {\em Transportation Research Part C: Emerging Technologies}, {\bf 124}, 102947.

\bibitem[\protect\citename{Reck {\em et~al.}, }2022]{reck2022mode}
Reck, Daniel~J, Martin, Henry, \& Axhausen, Kay~W. 2022.
\newblock Mode choice, substitution patterns and environmental impacts of shared and personal micro-mobility.
\newblock {\em Transportation Research Part D: Transport and Environment}, {\bf 102}, 103134.

\bibitem[\protect\citename{Rose \& Bliemer, }2009]{rose2009constructing}
Rose, John~M, \& Bliemer, Michiel~CJ. 2009.
\newblock Constructing efficient stated choice experimental designs.
\newblock {\em Transport reviews}, {\bf 29}(5), 587--617.

\bibitem[\protect\citename{Shaheen {\em et~al.}, }2013]{shaheen2013public}
Shaheen, Susan~A, Martin, Elliot~W, \& Cohen, Adam~P. 2013.
\newblock Public Bikesharing and Modal Shift Behavior: A Comparative Study of Early Bikesharing Systems in North America.
\newblock {\em International Journal of Transportation}, {\bf 1}(1).

\bibitem[\protect\citename{Shen, }2014]{shen2014latent}
Shen, Junyi. 2014.
\newblock Latent class model or mixed logit model? A comparison by transport mode choice data.
\newblock {\em Pages  127--136 of:} {\em The Applied Economics of Transport}.
\newblock Routledge.

\bibitem[\protect\citename{Sinha {\em et~al.}, }2021]{sinha2021practitioner}
Sinha, Pratik, Calfee, Carolyn~S, \& Delucchi, Kevin~L. 2021.
\newblock Practitioner’s guide to latent class analysis: methodological considerations and common pitfalls.
\newblock {\em Critical care medicine}, {\bf 49}(1), e63--e79.

\bibitem[\protect\citename{Tiwari, }2019]{tiwari2019micro}
Tiwari, A. 2019.
\newblock Micro-mobility: the next wave of urban transportation in India.
\newblock {\em YS Journal, January}.

\bibitem[\protect\citename{van Kuijk {\em et~al.}, }2022]{van2022preferences}
van Kuijk, Roy~J, de~Almeida~Correia, Gon{\c{c}}alo~Homem, van Oort, Niels, \& Van~Arem, Bart. 2022.
\newblock Preferences for first and last mile shared mobility between stops and activity locations: A case study of local public transport users in Utrecht, the Netherlands.
\newblock {\em Transportation Research Part A: Policy and Practice}, {\bf 166}, 285--306.

\bibitem[\protect\citename{van Mil {\em et~al.}, }2021]{van2021insights}
van Mil, Joeri~FP, Leferink, Tessa~S, Annema, Jan~Anne, \& van Oort, Niels. 2021.
\newblock Insights into factors affecting the combined bicycle-transit mode.
\newblock {\em Public Transport}, {\bf 13}(3), 649--673.

\bibitem[\protect\citename{Wang {\em et~al.}, }2023]{wang2023travel}
Wang, Kailai, Qian, Xiaodong, Fitch, Dillon~Taylor, Lee, Yongsung, Malik, Jai, \& Circella, Giovanni. 2023.
\newblock What travel modes do shared e-scooters displace? A review of recent research findings.
\newblock {\em Transport Reviews}, {\bf 43}(1), 5--31.

\bibitem[\protect\citename{Zhou {\em et~al.}, }2020]{zhou2020analysing}
Zhou, Heng, Norman, Richard, Xia, Jianhong~Cecilia, Hughes, Brett, Kelobonye, Keone, Nikolova, Gabi, \& Falkmer, Torbjorn. 2020.
\newblock Analysing travel mode and airline choice using latent class modelling: A case study in Western Australia.
\newblock {\em Transportation Research Part A: Policy and Practice}, {\bf 137}, 187--205.

\end{thebibliography}

\end{sloppypar}
\end{small}


\end{document}